\documentclass[review,author-year]{elsarticle}

\usepackage{amsmath,amsthm,amssymb}
\usepackage{mathtools}
\usepackage{graphicx}
\usepackage{xcolor}
\usepackage{longtable}
\usepackage{multirow}
\usepackage{subcaption}
\usepackage{float}
\usepackage{setspace}
\usepackage{hyperref}
\newcommand{\md}[2]{\frac{\mathrm{D} #1}{\mathrm{D} #2}}
\newcommand{\od}[2]{\frac{\mathrm{d} #1}{\mathrm{d} #2}}
\newcommand{\odtwo}[2]{\frac{\mathrm{d}^2 #1}{\mathrm{d} #2 ^2}}
\newcommand{\pd}[2]{\frac{\partial #1}{\partial #2}}

\newcommand{\lFeO}{^l_{Fe,O}}
\newcommand{\sFe}{^s_{Fe}}
\newcommand{\slFe}{^{s,l}_{Fe}}
\newcommand{\slFeO}{^{s,l}_{Fe,O}}

\journal{EPSL}

\begin{document}
	
	\begin{frontmatter}
		
		\title{A regime diagram for the slurry F-layer at the base of Earth's outer core\tnoteref{mytitlenote}}
		\author{Jenny Wong\fnref{address1,address2}\corref{correspondingauthor}}
		\cortext[correspondingauthor]{Corresponding author}
		\ead{wong@ipgp.fr}
		\author{Christopher J. Davies \fnref{address3}}
		\author{Christopher A. Jones \fnref{address4}}
		
		\address[address1]{Universit\'{e} de Paris, Institut de physique du globe de Paris, CNRS, F-75005 Paris, France}
		\address[address2]{EPSRC Centre for Doctoral Training in Fluid Dynamics, University of Leeds, Leeds, LS2 9JT, UK}
		\address[address3]{School of Earth and Environment, University of Leeds, Leeds, LS2 9JT, UK}
		\address[address4]{School of Mathematics, University of Leeds, Leeds, LS2 9JT, UK}
		
		\begin{abstract} % 300 word limit
			Seismic observations of a slowdown in P wave velocity at the base of Earth’s outer core suggest the presence of a stably-stratified region known as the F-layer. This raises an important question: how can light elements that drive the geodynamo pass through the stably-stratified layer without disturbing it? We consider the F-layer as a slurry containing solid particles dispersed within the liquid iron alloy that snow under gravity towards the inner core. We present a regime diagram showing how the dynamics of the slurry F-layer change upon varying the key parameters: P\'{e}clet number ($Pe$), the ratio between advection and chemical diffusion; Stefan number ($St$), the ratio between sensible and latent heat; and Lewis number ($Le$), the ratio between thermal and chemical diffusivity. We obtain four regimes corresponding to stable, partially stable, unstable and no slurries. No slurry is found when the heat flow at the base of the layer exceeds the heat flow at the top, while a stably-stratified slurry arises when advection overcomes thermal diffusion ($Pe \gtrsim Le$) that exists over a wide range of parameters relevant to the Earth's core. Our results estimate that a stably-stratified F-layer gives a maximum inner-core boundary (ICB) body wave density jump of $\Delta \rho_\textrm{bod} \leq 534 \  \mathrm{kg} \mathrm{m}^{-3}$ which is compatible with the lower end of the seismic observations where $280 \leq \Delta \rho_\textrm{bod} \leq 1,100  \  \mathrm{kg} \mathrm{m}^{-3}$ is reported in the literature. With high thermal conductivity the model predicts an inner core age between $0.6$ and $1.2 \ \mathrm{Ga}$, which is consistent with other core evolution models. Our results suggest that a slurry model with high core conductivity predicts geophysical properties of the F-layer and core that are consistent with independent seismic and geodynamic calculations.
		\end{abstract}
		
		\begin{keyword}
			slurry, iron snow, inner core, crystallisation, F-layer
		\end{keyword}
		
	\end{frontmatter}
	
%	\linenumbers
	
	\section{Introduction}
	% Impactful introduction
	Seismic, geomagnetic and dynamical explanations for and against the presence of stably-stratified layers in the Earth's liquid core is an active research topic of significant geophysical interest. Confirmation of their existence would warrant a shift away from the usual approximation that the core is adiabatically stratified with thin boundary layers. Stably-stratified layers should exhibit dynamics that distinguish them from the turbulent bulk of the core; elucidating this behaviour will help to uncover their signature in seismic and geomagnetic observations \cite{Hardy2019}. In this paper we focus on the F-layer at the base of the liquid core to further understand the conditions where stable stratification can be sustained.
	
	% F-layer
	There is a strong consensus that a slowdown in the P wave velocity compared with the Preliminary Reference Earth Model (PREM) \cite{Dziewonski81} is observed at the base of the Earth's outer core \cite{Souriau91, Ohtaki15, Zou08}. PREM follows the Adams-Williamson equation that assumes the outer core is adiabatically stratified and homogeneous throughout, therefore a P wave velocity lower than PREM is attributed to an anomalously higher density structure than expected. This departure in density away from neutral stability means that the seismic observations suggest a stably-stratified layer exists that cannot be explained by adiabatic compression alone. Estimates of the layer thickness vary from $150 \ \mathrm{km}$ \cite{Souriau91} to $400 \ \mathrm{km}$ \cite{Ohtaki15} thick, which greatly exceeds the thermal diffusion length scale, hence the layer cannot be simply explained by a thermal boundary layer alone and another mechanism is required to maintain stratification \cite{Deguen12}.
	
	% Why core crystallisation is important
	The F-layer is intimately linked to the geodynamo process that generates Earth's magnetic field. Motion of the liquid iron core is powered by heat extracted at the core-mantle boundary (CMB), which leads to freezing of the solid inner core from the centre of the Earth because the melting curve is steeper than the core adiabat. Freezing releases latent heat and light elements into the liquid; light elements drive compositional convection in the core and is thought to be the main power source for the dynamo at the present day \citep{Nimmo15a}. The key issue is buoyant light material excluded from the inner core must pass through the F-layer and into the overlying core while preserving stable stratification.
	
	% Geodynamic models
	Previous geodynamic models try to explain the F-layer by inner core translation \cite{Alboussiere12}, a thermochemical layer \cite{Gubbins08}, or a slurry (iron snow) layer \cite{Loper77, Wong18a}. Inner core translation supposes melting occurs on the eastern hemisphere of the inner core while freezing occurs on the western hemisphere to form a dense layer above the inner-core boundary (ICB), though recent upward revisions of the thermal conductivity of iron \cite{deKoker12,Pozzo12} together with the likely presence of compositionally stabilising conditions suggest that the instability cannot arise in the present-day \cite{Pozzo14,Deguen18}. \cite{Gubbins08} proposed a thermochemical model with the F-layer on the liquidus that succeeded in producing a stably-stratified layer, however, the model effectively imposed a stable composition without a physical mechanism explaining why. Wong \textit{et al.} \citep{Wong18a} explains this mechanism in a self-consistent way by proposing a slurry layer that captures the physics of how light element and solid is transported. The authors used a simple box model to demonstrate that a slurry layer produced stably-stratified layers that match the seismic observations of density and layer thickness. In this work we focus on the slurry scenario and build upon Wong \textit{et al.} \citep{Wong18a}, hereafter referred to as W18.
	
\begin{figure}
		\centering
		\includegraphics[width=\textwidth]{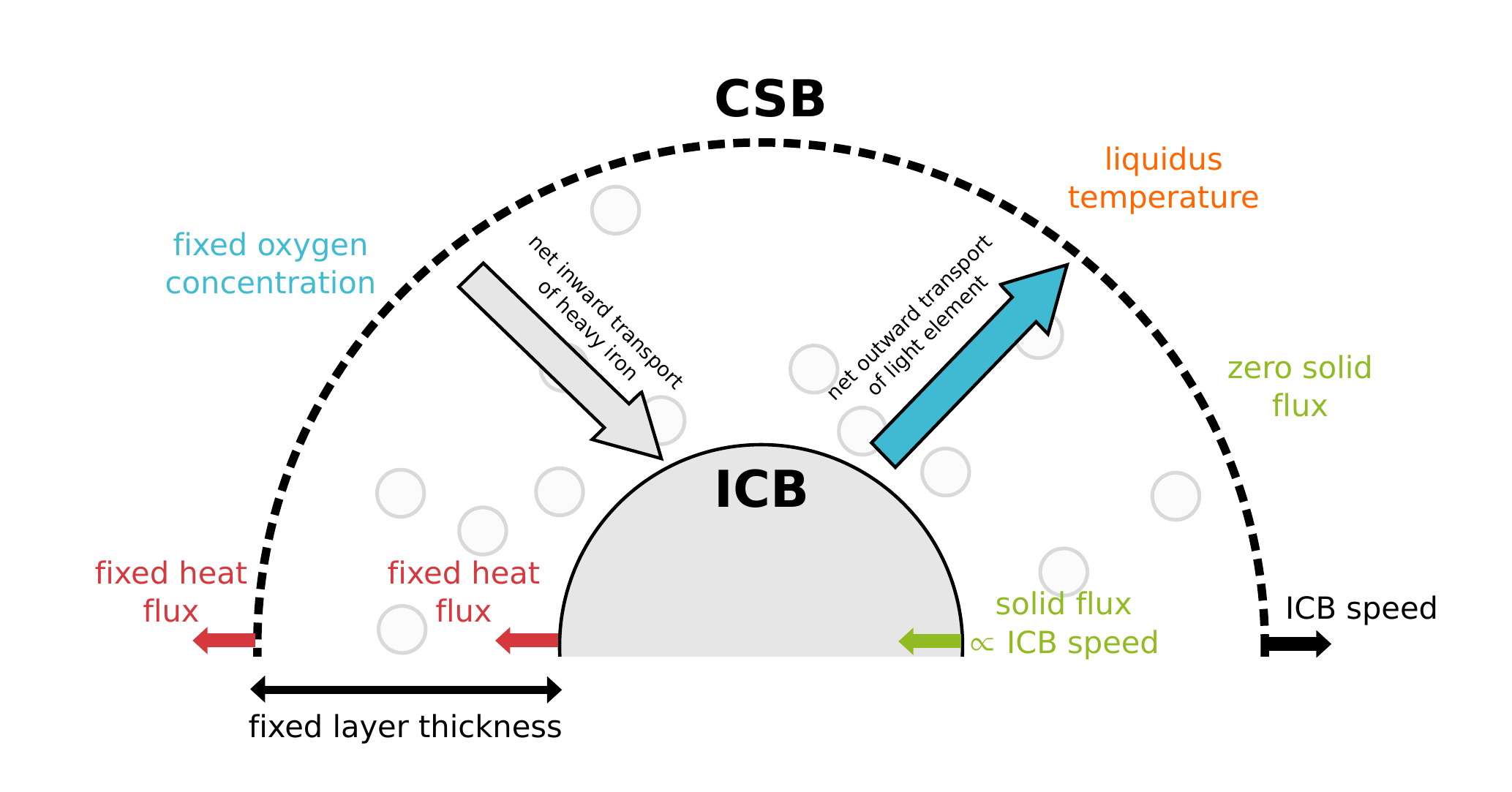}
		\caption{A sketch of the slurry layer at the base of Earth's outer core, with the boundary conditions imposed at the inner core boundary (ICB) and core slurry boundary (CSB) described in Section \ref{sec:mod}.}
		\label{fig:schematic}
	\end{figure}

	% Slurry model
	In the slurry, pure solid iron particles crystallise throughout the entire F-layer while light elements remain in the liquid (see Figure \ref{fig:schematic} for a sketch). Heavy grains of iron fall under gravity to accumulate at the base of the layer, thereby producing a net inward transport of dense solid and a net outward transport of light elements to give an overall stable density stratification. W18 simplified the full slurry theory by Loper and Roberts \citep{Loper77, Loper87} and developed a reduced, thermodynamically self-consistent framework that accounts for solid and liquid phase, in addition to the influence of pressure, temperature and composition. The fraction of solid in a slurry is assumed to be small, and so cannot transmit shear waves to create an impedance contrast at the top of the F-layer that would otherwise have been seismically observable.
	
	% Model assumptions
	The principal assumptions of the W18 slurry model are (1) the fast-melting limit and (2) a binary alloy. (1) supposes that an infinitesimal material volume contains either solid or liquid phase exclusively so that the system is in phase equilibrium. As a consequence, minimising the Gibbs free energy constrains the slurry temperature to the liquidus. (2) presumes an Fe-O composition since oxygen is able to explain the core density deficit \cite{Birch52}. \textit{Ab initio} calculations show that oxygen almost entirely partitions into the liquid phase upon freezing \cite{Alfe02c} and cannot occur  with silicon (another potential light element candidate) in large quantities \cite{Badro14}. The solid produced by the slurry is thereby assumed to be composed of pure iron, which avoids a complicated particle history dependence that would be difficult to model, where the growth of solid grains at different pressure-temperature conditions affects the composition of light elements in the solid.
	
	% Model differences to Wong et al. (2018)
	We generalise the W18 model by moving from a Cartesian to spherical geometry and update the boundary conditions to a more geophysically realistic setup. We suppose that the inner core is isothermal and not convecting due to the upward revisions of the thermal conductivity \cite{Gomi13,Williams18} of iron alloys at core conditions, which means that the present-day inner core is unlikely to convect.	We consider a thin compacting layer on the order of kilometres thick on the solid side of the inner core \cite{Deguen07}, parameterised by interfacial freezing at the ICB which generates a flux of latent heat into the slurry. Accordingly, the ICB advances at a rate composed of the interfacial freezing speed and the snow speed given by the accumulation of solid iron particles from the slurry. Our model self-consistently determines the ICB speed and therefore allows the inner core growth rate to be independently determined, whereas this was fixed in W18 and yielded unrealistically high ICB heat flows. Following the fast-melting limit, we set the temperature at the top of the layer to the liquidus temperature for the bulk core composition obtained from the literature, so that the ICB temperature is free to be self-determined by the slurry whereas this was fixed in W18.
	
	% Paper outline
	In this paper, we perform a systematic parameter search to elucidate the conditions that promote stable stratification of the F-layer. The generalised model of W18 and the derivation of the dimensionless system with its associated dimensionless control parameters is given in Section \ref{sec:mod}. In Section \ref{sec:res}, we present and discuss example solutions of the temperature, oxygen concentration, solid flux and density given by the slurry model with different layer thicknesses. We proceed to test the sensitivity of the model to the boundary conditions on the temperature and composition at the top of the layer, before mapping a regime diagram of the slurry based on the main dimensionless control parameters. This vastly improves upon the computed solution space of W18 and uncovers the possible regimes occupied by the slurry. We demonstrate the predictive power of our model for seismic observations and inner core age, before concluding with a summary of our findings in Section \ref{sec:con}.
	
	\section{Model and Methods} \label{sec:mod}
	\subsection{Dimensional equations}
	Detailed development of the slurry model is given W18 so here we present the key details. We start with the general equations for the conservation of oxygen, temperature and the liquidus constraint from W18 (equations (4),(6) and (13)), given by
	\begin{align}
		\rho^{sl} \md{\xi}{t} &= - \nabla \cdot \mathbf{i}, \label{eqn:gen_xi} \\ 
		\rho^{sl} c_p \md{T}{t} & = \nabla \cdot \left(k \nabla T + L \mathbf{j}\right) + \rho^{sl} L \md{\phi}{t}, \label{eqn:ene} \\
		\nabla T & = \frac{T \Delta V \slFe}{L} \nabla p - \frac{T \xi^l \left(\partial \mu / \partial \xi^l \right)}{L} \nabla \xi^l, \label{eqn:gen_liq}
	\end{align}
	where $p$ is the pressure, $T$ is the temperature, $\xi$ is the oxygen concentration, $\xi^l$ is the oxygen concentration in the liquid phase, $\phi$ is the solid fraction, $\rho^{sl}$ is the reference density of the slurry taken to be the PREM value at the CSB, $\mathbf{i}$ is the oxygen flux vector, $\mathbf{j}$ is the solid flux vector, $c_p$ is the specific heat capacity, $k$ is the thermal conductivity, $L$ is the latent heat of fusion, $\Delta V \slFe$ is the change in specific volume between liquid iron and solid iron, and $\partial \mu / \partial \xi^l$ is the thermodynamic derivative of the chemical potential, $\mu$, with respect to $\xi^l$. Throughout this paper superscripts $s$, $l$ and $sl$ denote solid, liquid or slurry phases respectively, and subscripts $Fe$ and $O$ denote the iron and oxygen components, respectively. For further reference, a table of symbols and values is provided in \ref{sup:val}, Table \ref{tbl:val}.
	
	We adopt a spherical, one-dimensional geometry where we assume a global F-layer in a non-convective steady state that depends on radius only and excludes lateral variations. The layer thickness, $d$, is fixed so that the slurry is defined over the interval $[r^i,r^{sl}]$, where $r^i$ denotes the ICB radius and $r^{sl} \equiv r^i + d$ is the core-slurry boundary (CSB). We are interested in a timescale that is long compared with the timescale of freezing and comparable to the evolution of the inner core \cite{Gubbins03}, so we assume a Boussinesq slurry with a reference state in hydrostatic equilibrium, and resolve the slurry layer in a frame moving at the rate of inner core growth, $v\mathbf{\hat{r}}$, with $v$ the ICB speed and $\mathbf{\hat{r}}$ the unit vector pointing outwards and normal to the inner core surface. Mass conservation $\nabla \cdot \mathbf{v} =0$ implies that $v(r) = v^i (r^i/r)^2$, where the changes in ICB speed due to radius are negligible and we therefore assume that $v$ is constant. This speed is composed of two parts, $v = v_s + v_f$, where $v_s$ is the snow speed, in which all solid particles from the slurry are assumed to accumulate at the base of the layer; and $v_f$ is the freezing speed, which represents the growth of the inner core due to compaction on the solid side of the ICB, and assumed to occur quickly compared with the slurry timescale. The material derivative is hence $\mathrm{D} / \mathrm{D} t \longrightarrow -v \ \mathrm{d} / \mathrm{d} r$. As in W18, the fraction of solid, $\phi$, in the slurry is small so that $\phi \ll 1$, and we assume that their variations, $\mathrm{d}\phi$, are negligible (see discussion in \ref{sec:dphi}). We apply the above assumptions to equations (\ref{eqn:gen_xi} -- \ref{eqn:gen_liq}), to obtain, 
	\begin{align}
	\phantom{i + j + k}
	&\begin{aligned}
	\mathllap{- v \rho^{sl} \od{\xi^l}{r}} &= \frac{1}{r^2} \od{}{r} \left(r^2 \frac{\rho ^{sl} \Delta V\slFeO D_O}{RT \frac{1000}{a_O}} \exp {\left(\frac{F \left(r-r^i \right)}{d}\right)} \od{p}{r} \right)\\
	&\qquad \qquad \qquad \qquad \qquad \qquad + \xi^l \od{j}{r} + j \od{\xi^l}{r} + \frac{2}{r} \xi^l j, \label{eqn:dXi}
	\end{aligned}\\
	&\begin{aligned}
	\mathllap{- v \rho ^{sl} c_p \od{T}{r}} &= k\odtwo{T}{r}  + \frac{2}{r} k \od{T}{r} + L \od{j}{r} + \frac{2}{r} L j , \label{eqn:dTemp}
	\end{aligned} \\
	&\begin{aligned}
	\mathllap{\od{T}{r}} &= - \frac{T \Delta V\slFe }{L} g \rho - \frac{R T^2 1000}{a_O L} \od{\xi^l}{r}\label{eqn:dLiq}.
	\end{aligned}
	\end{align}
	
	\bigskip
	
	On the LHS of (\ref{eqn:dXi}), we have the advection of oxygen, and on the RHS, the first term is the effect of barodiffusion and the last three terms describe the physical displacement of oxygen as solid iron particles sediment. On the LHS of (\ref{eqn:dTemp}), we have the advection of heat, and on the RHS, the first two terms correspond to thermal diffusion and the last two terms correspond to the latent heat release associated with phase change. The liquidus constraint (\ref{eqn:dLiq}) describes the change in the liquidus temperature as a consequence of changes in pressure, given by the first term on the RHS, and changes in composition, given by the second term on the RHS.
	
	We have applied ideal solution theory \cite{Gubbins04b} so that the thermodynamic derivative of the chemical potential with respect to $\xi^l$ can be expressed as $\xi^l \partial \mu / \partial \xi^l = RT 1000 / a_O$, where $R$ is the gas constant and $a_O$ is the atomic weight of oxygen. We expect penetrative convection to occur at the top of the layer because of the velocity difference between the non-convective slurry and the overlying convective outer core. We have thus invoked a turbulent mixing sublayer below the CSB that promotes the transport of oxygen out of the layer and also allows us to impose a vanishing solid flux boundary condition at the CSB (see equation (\ref{eqn:djCSB}) below). This effect is modelled by modifying the self-diffusion coefficient of oxygen to take the exponential form
	\begin{align}
	\bar{D} = D_O \exp{\left(\frac{F \left(r-r^i \right)}{d}\right)}, \label{eqn:dSelf}
	\end{align}
	where $F$ is the mixing parameter that appears in the first term on the RHS of (\ref{eqn:dXi}), and $D_O$ is the self-diffusion coefficient of oxygen taken from the literature \cite{Pozzo13}.
	
	\subsection{Boundary conditions}
	We have three equations (\ref{eqn:dXi}),(\ref{eqn:dTemp}), and (\ref{eqn:dLiq}) for three output variables $T, \xi^l, j$, and two output parameters (eigenvalues) $F$ and $v$. The system is fourth order, therefore we require six boundary conditions to determine a unique solution. We suppose that the oxygen concentration is constant at the CSB and equal to the uniform value in the bulk of the core, $\xi^{sl}$, since the liquid core is vigorously convecting and its composition does not change much over the timescales considered \cite{Davies15a}, so that
	\begin{align}
	\xi^l(r^{sl}) &= \xi^{sl}. \label{eqn:dxiCSB}
	\end{align}
	At the ICB the solid flux is proportional to the ICB speed and at the CSB the solid flux vanishes, so that
	\begin{align}
	j(r^i) &= - \rho^s_- v, \label{eqn:djICB} \\
	j(r^{sl}) &= 0, \label{eqn:djCSB}
	\end{align}
	where $\rho^s_-$ denotes the density on the solid side of the ICB.
	
	We fix the heat flux per unit area at the CSB to a constant value, $q^{sl}$, and this is a parameter to be varied since there are no independent estimates. By Fourier's law, the boundary condition on the CSB temperature gradient is
	\begin{align}
	\od{T}{r} \bigg \rvert_{r=r^{sl}} &= -\frac{q^{sl}}{k}. \label{eqn:dtempgradCSB}
	\end{align}
	In the following two boundary conditions, we digress from the conditions imposed in W18. First, compaction below the F-layer releases latent heat at the interface while there is no specific heat contribution from an isothermal IC, whereas W18 assumed specific heat loss from the inner core. Second, we assume that the temperature at the top of the layer is coincident with the liquidus temperature of the bulk composition, $T^{sl}(r^{sl})$, known from estimates given in the literature \cite{Pozzo13,Davies15b}, so that
	\begin{align}
	\od{T}{r} \bigg \rvert_{r=r^i} &= - \frac{\rho^s_- v_f L}{k} = -\frac{q^s}{k }, \label{eqn:dtempgradICB} \\
	T(r^{sl}) &= T^{sl} (r^{sl}). \label{eqn:dtempCSB}
	\end{align}
	where $q^s$ is the ICB heat flux per unit area.
	
	\subsection{Dimensionless equations}
	We derive the dimensionless equations using the following scalings,
		\begin{align}
		r&=r^{sl}\hat{r}, & T &=\frac{q^{sl}}{\rho^{sl} c_p v_f } \hat{T}, \nonumber \\
		\xi^l&=\xi^{sl} \hat{\xi},  & j &= \rho^s_- v_f \hat{\jmath}, \nonumber \\
		g &= g^{sl} \hat{g}, & \rho &= \rho^{sl} \hat \rho, \nonumber \\
		v &= v_f \hat{v}
		\label{eqn:scalings}
		\end{align}
	where the hat symbol denotes a dimensionless quantity that depends on radius. The slurry layer is defined over the dimensionless interval $[r^i/r^{sl},1]$. Equations (\ref{eqn:dXi})--(\ref{eqn:dLiq}) respectively become
		\begin{align}
		\phantom{i + j + k }
		&\begin{aligned}
		\mathllap{-\hat{v}\od{\hat{\xi}}{\hat{r}}} &= - \frac{1}{\hat{r}^2} \od{}{\hat{r}} \left( \frac{Li_p R_\rho}{Li_\xi Pe St R_v} \frac{\hat{g} \hat{\rho} \hat{r}^2}{\hat{T}}  \exp\left[\frac{F\left(r^{sl}\hat{r} - r^i \right)}{d}\right] \right) \\
		&\qquad \qquad \qquad \qquad \qquad \qquad \qquad + \hat{\xi} \od{\hat{\jmath}}{\hat{r}} + \hat{\jmath} \od{\hat{\xi}}{\hat{r}} + \frac{2}{\hat{r}}\hat{\xi}  \hat{\jmath}, \label{eqn:nXi}
		\end{aligned}\\
		&\begin{aligned}
		\mathllap{-\hat{v} \od{\hat{T}}{\hat{r}}} &= \frac{Le}{Pe} \left(\odtwo{\hat{T}}{\hat{r}} + \frac{2}{\hat{r}}\od{\hat{T}}{\hat{r}} \right) + \frac{1}{St} \left(\od{\hat{\jmath}}{\hat{r}} +\frac{2}{\hat{r}}\hat{\jmath} \right) , \label{eqn:nTemp}
		\end{aligned} \\
		&\begin{aligned}
		\mathllap{\od{\hat{T}}{\hat{r}}} &= -Li_p \hat{g} \hat{\rho} \hat{T} - \frac{Li_\xi St}{R_\rho} \hat{T}^2 \od{\hat{\xi}}{\hat{r}} \label{eqn:nLiq}.
		\end{aligned}
		\end{align}
	where the dimensionless numbers are defined as
		\begin{gather}
		R_\rho = \frac{\rho^{sl}}{\rho^s_-}, \quad
		R_v = \frac{\Delta V \slFe}{\Delta V \slFeO}, \quad
		Li_p  \equiv \frac{\Delta V\slFe g^{sl} \rho^{sl} r^{sl}}{L}, \quad
		Li_\xi  \equiv \frac{1000 R \xi^{sl}}{a_O c_p}, \nonumber \\
		Pe  \equiv \frac{v_f r^{sl}}{D_O}, \quad
		St  \equiv \frac{q^{sl}}{ \rho ^{s}_- v_f L}, \quad
		Le  \equiv \frac{k}{\rho^{sl} c_p D_O}.
		\end{gather}
	$R_\rho$ is the ratio between the reference density and the density on the solid side of the ICB, and $R_v$ is the ratio between the change in specific volumes upon phase change of pure iron and the iron alloy. The dimensionless numbers $Li_p$ and $Li_\xi$ arise from the pressure and compositional parts of the liquidus constraint (\ref{eqn:nLiq}), respectively. $Pe$ is the P\'{e}clet number that measures the ratio between advection and chemical diffusion. $St$ is the Stefan number and gives the ratio between sensible and latent heat. $Le$ is the Lewis number that describes the ratio between thermal and chemical diffusivity.
	
	Inserting the scalings (\ref{eqn:scalings}) into (\ref{eqn:dxiCSB})--(\ref{eqn:dtempCSB}) yields the dimensionless boundary conditions
	\begin{align}
	\hat{T}(1) &= \frac{T^{sl} c_p R_\rho}{St L}, \label{eqn:nondim_tempCSB} \\
	\od{\hat{T}}{\hat{r}} \bigg \rvert_{\hat{r}=\frac{r^i}{r^{sl}}} &= -\frac{Pe}{St Le}, \label{eqn:nondim_tempgradICB} \\
	\od{\hat{T}}{\hat{r}} \bigg \rvert_{\hat{r}=1} &= -\frac{Pe}{Le}, \label{eqn:nondim_tempgradCSB} \\
	\hat{\xi}(1) &= 1, \\
	\hat{\jmath}\left(\frac{r^i}{r^{sl}}\right) &= -\hat{v}, \\
	\hat{\jmath}(1) &= 0. \label{eqn:nondim_jCSB}
	\end{align}
	
	\subsection{Model parameters and constraints} \label{sec:met}
	We solve the dimensionless system numerically using \texttt{solve\_bvp}, which is a boundary value problem solver included in \texttt{scipy} -- an open source Python library \cite{Virtanen2020}. We rewrite the problem as a system of first-order ODEs, and vary the model parameters $Pe$, $St$, $Le$, $Li_p$, $Li_\xi$, $R_\rho$, $R_v$ and $T^{sl}$ (see Table \ref{tbl:param}). The numerical code written to solve the slurry equations is freely available online \footnote{\url{https://github.com/jnywong/nondim-slurry}}. Unless otherwise specified, all other physical properties of the F-layer are assumed constant, as specified in Table \ref{tbl:val}.
	
	$Pe$ varies because of the range of seismic estimates of the F-layer thickness, $d$, and the unknown freezing speed, $v_f$. We obtain a proxy for $v_f$ through the ICB heat flow, $Q^s$, which we presume cannot be much greater than the adiabatic value estimated as $Q^a = 1.6 \ \mathrm{TW}$ \cite{Pozzo14}, though we allow a wider margin of $Q^s$ up to $3.2 \ \mathrm{TW}$ to account for uncertainties also inherent in the estimate of $Q^a$ in the literature. $St$ varies due to $Q^s$ and $Q^{sl}$. While there is no geophysical constraint on $Q^{sl}$, we test $Q^{sl}$ over a wide range up to $12 \ \mathrm{TW}$ which is verified \textit{a posteriori} to give $5 \leq Q^c \leq 17 \ \mathrm{TW}$ \cite{Nimmo15a,Lay08} (see \ref{sec:cmb} for the calculation of $Q^c$). $Le$ depends on the thermal conductivity and we explore two values that represent the higher and the lower estimates given in the literature \cite{Pozzo12,Williams18,Stacey07a}. $Li_p$ and $R_\rho$ change according to the layer thickness and also the reference density, $\rho^{sl}$, though the density on the solid side of the ICB remains the same and is taken from PREM, $\rho_-^s = 12,764 \ \mathrm{kg} \mathrm{m}^{-3}$. $Li_\xi$ depends on $\xi^{sl}$, wherein the precise amount of oxygen present in the bulk of the Earth's core is difficult to constrain with reported values varying between $0.1$ and $11.0 \ \textrm{mol.\%}$ for core chemistry models that include other light elements in addition to oxygen \cite{Hirose13}. Varying $\xi^{sl}$ between $2.0$ and $12.0 \ \textrm{mol.}\%$ changes $R_v$, which depends on both the layer thickness and the CSB oxygen concentration and in turn affects the change in specific volume between solid iron and the liquid iron alloy. % with intervals of $0.5 \ \textrm{mol.}\%$.
	
	\begin{table}[ht]
		\centering
		\begin{tabular}{p{5.25cm}llp{3.75cm}} \hline
			Input parameter      & Symbol & Units                                           & F-layer                  \\ \hline
			Layer thickness      & $d$    & $\mathrm{km}$                                   & $150,200,250,300,350,400$           \\
			Thermal conductivity & $k$    & $\mathrm{W} \ \mathrm{m}^{-1} \mathrm{K}^{-1}$  & $30, 100$            \\
			ICB heat flux        & $Q^s$  & $\mathrm{TW}$                                   & $0$ -- $3.20$            \\
			CSB heat flux        & $Q^{sl}$ & $\mathrm{TW}$                                 & $0$ -- $12$           \\
			Sedimentation prefactor & $k_\phi$ & $\mathrm{kg} \mathrm{m}^{-3} \mathrm{s}$ & $10^{-5}$ -- $10^{-1}$     \\
			CSB liquidus temperature  & $T^{sl}$ & $\mathrm{K}$                             & $4,500$ -- $6,000$         \\
			CSB oxygen concentration  & $\xi^{sl}$ & $\mathrm{mol.} \%$                     & $2$ -- $12$              \\ \\
			P\'{e}clet number    & $Pe$   &                                                 & $0$ -- $2500$            \\
			Stefan number        & $St$   &                                                 & $0$ -- $3$               \\
			Lewis number         & $Le$   &                                                 & $354$ -- $1196$          \\
			Liquidus number %
			(pressure)           & $Li_p$ &                                                 & $0.163$ -- $0.220$       \\
			Liquidus number %
			(compositional)      & $Li_\xi$ &                                               & $0.004$ -- $0.028$       \\
			Ratio between solid %
			and reference density & $R_\rho$ &                                              & $0.940$ -- $0.952$       \\
			Ratio between change %
			in specific volumes %
			upon phase change of %
			pure iron and iron %
			alloy                & $R_v$ &                                                  & $0.203$ -- $0.235$ \\
			\hline
		\end{tabular} %}
		\caption{Relevant dimensional and dimensionless parameter range for the F-layer, taking other physical parameters as constant given in Table \ref{tbl:val}.}
		\label{tbl:param}
	\end{table}
	
	We constrain $T^{sl}$ in boundary condition (\ref{eqn:nondim_tempCSB}) from the melting curves of iron, which is usually reported in the range of $5,500 \ \mathrm{K}$ \cite{Sinmyo19} to $6,350 \ \mathrm{K}$ \cite{Jackson13}. Light elements present in the iron alloy depress the liquidus temperature of pure iron, and a typical value of $\Delta T_\xi = 700 \ \mathrm{K}$ is used, though this can also vary between $500$ and $1,000 \ \mathrm{K}$ \cite{Hirose13}. We therefore opt to vary $T^{sl}$ between $4,500$ and $6,000 \ \mathrm{K}$. 
	
	We constrain the results of the parameter study in three ways:
	\begin{enumerate}
		\item by the seismically determined density jump at the ICB \cite{Gubbins08}. The jump obtained from normal modes, $\Delta \rho_\textrm{mod}$, has a long wavelength on the order of hundreds of kilometres and represents the difference between the average densities of the top of the inner core and the bottom of the outer core. The jump obtained from body waves, $\Delta \rho_\textrm{bod}$, has a short wavelength on the order of several kilometres, and therefore represents the difference in densities either side of the ICB. The difference between the normal mode and body wave estimates therefore points to a density anomaly caused by the F-layer. Normal mode studies suggest $600 \ \mathrm{kg} \ \mathrm{m}^{-3} \leq \Delta \rho_\textrm{mod} \leq 820 \pm 180 \  \mathrm{kg} \ \mathrm{m}^{-3}$ \cite{Dziewonski81,Masters03}, whereas body wave studies suggest $520 \pm 240 \ \mathrm{kg} \ \mathrm{m}^{-3} \leq \Delta \rho_\textrm{bod} \leq 1,100 \ \mathrm{kg} \ \mathrm{m}^{-3}$ \cite{Koper05,Tkalcic09}. Hence solutions obtained from the slurry model should satisfy a maximum density jump across the layer of $\mathrm{max}(\Delta \rho_\text{mod} - \Delta \rho_\text{bod}) = 1,000 - 280 = 720 \ \mathrm{kg} \ \mathrm{m}^{-3}$, and a minimum bound $\mathrm{min}(\Delta \rho_\text{mod} - \Delta \rho_\text{bod}) <0$ is not specified since the seismic observations from different studies vary in their approaches, so we consider zero as the lower bound.
		\item solutions should be consistent with estimates of the present-day CMB heat flux, which should be between $5$ and $17 \ \mathrm{TW}$ \cite{Nimmo15a,Lay08}
		\item we assess the stability of the layer and determine whether the solution is stable, partially stable or unstable, and reject solutions that are unstable.
	\end{enumerate}
	
	To apply the constraints on density we calculate the density jump across the slurry layer, $\rho_+^s - \rho^{sl}$, where $\rho_+^s$ is the density on the slurry side of the ICB. The total density is
	\begin{align}
	\rho = \rho_H + \rho',
	\end{align}
	where
	\begin{align}
	\rho_H = \left(\frac{g}{K}(r-r^{sl}) + \frac{1}{\rho^{sl} }\right)^{-1}, \label{eqn:den_hyd}
	\end{align}
	is the hydrostatic part, in which $K$ denotes the bulk modulus, and
	\begin{align}
	\rho' = \rho ^{sl} \left[-\alpha T' - \alpha_\xi {\xi^l}' + (\alpha_\phi+\alpha_\xi \xi^l) \phi' \right]. \label{eqn:den_flu} 
	\end{align}
	is the density perturbation, where $T'$, ${\xi^l}'$ and $\phi'$ are perturbations in the decomposition
		\begin{align*}
		T &= T^{sl} + T', \\
		\xi^l &= \xi^{sl} + {\xi^l}', \\
		\phi &= \phi^{sl} + \phi',
		\end{align*}
	with $\phi^{sl} = 0$ at the CSB. In the equation of state (\ref{eqn:den_flu}), the first two terms are well known from double diffusive/thermochemical convection theory, whereas the last term is unique to the slurry. The expansion coefficient of the solid is defined as
		\begin{equation*}
		\alpha_\phi = \rho ^{sl} \Delta V \slFeO,
		\end{equation*}
	(see derivation in \ref{sec:def}). To determine $\phi'$, the dimensional solid flux yielded from the solution of the slurry equations is related to the solid fraction by
	\begin{equation}
	\mathbf{j} = b(\phi) \Delta V \slFeO \nabla p.
	\end{equation}
	We employ a Stokes' flow model of mobility \cite{Loper80} where we assume that the solid iron particles created in the slurry layer drift towards the ICB under the influence of gravity. Consequently the sedimentation coefficient is given by
	\begin{align}
	b(\phi) = k_\phi \phi^{5/3} = \left(\frac{\rho^s_- (\rho^{sl})^2}{162 \pi^2 \nu^3 N^2}\right)^{1/3} \phi^{5/3}, \label{eqn:stokes}
	\end{align}
	where $N$ is the number of particles in a unit volume. We write $k_\phi$ to group the prefactors that multiply the solid fraction. This $k_\phi$ sedimentation prefactor introduces a degree of freedom since $N$ and the value of the kinematic viscosity, $\nu$, in the slurry is unknown. We further consider this issue in Section \ref{sec:sen}, where we conduct a sensitivity analysis on $k_\phi$.
	
	We evaluate the gradient of the density perturbations to assess the stability of the layer. Solving the slurry equations yields $T$, $\xi^l$ and $\phi$ and subtracting the reference values from these quantities gives the perturbations $T^\prime$, $\xi^\prime$ and $\phi^\prime$, which allows us to evaluate the density perturbation (\ref{eqn:den_flu}). If the layer is stable then the change in density relative to the hydrostatic reference should decrease as the radius increases, i.e. $\mathrm{d} \rho' /\mathrm{d}r < 0$, and vice versa if the layer is unstable, i.e. $\mathrm{d} \rho' /\mathrm{d}r > 0$. We relax the condition for stability slightly by stipulating that the layer is partially stable if $\mathrm{d} \rho' /\mathrm{d}r < 0$ for at least $100 \ \mathrm{km}$ of the layer since the seismic resolution from normal modes is on the order of this value.
	
	\section{Results and discussion} \label{sec:res}
	\subsection{Example solution} \label{sec:example}
	\begin{figure}[ht]
		\centering
		\includegraphics[width=\textwidth]{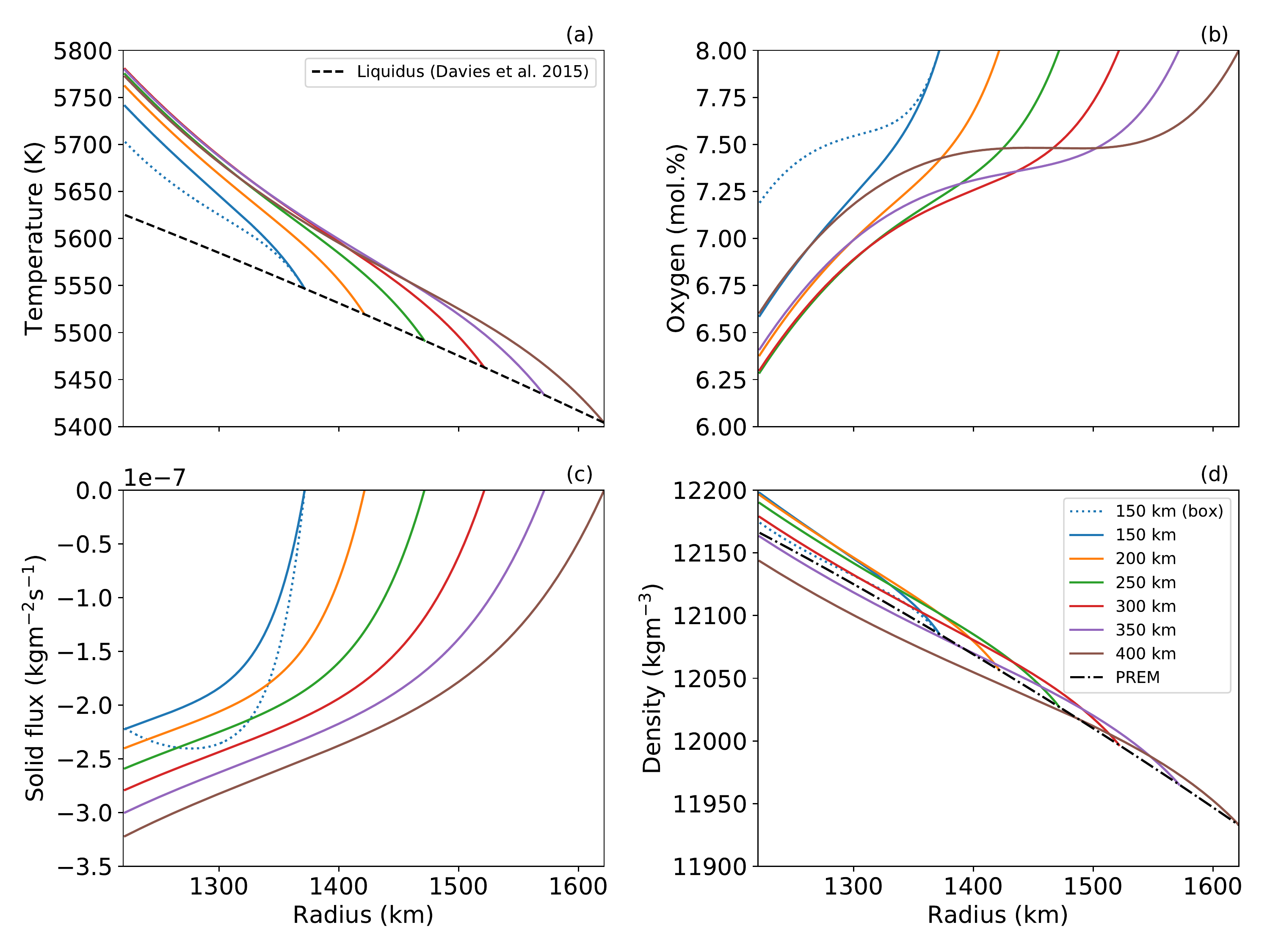}
		\caption{Profiles of (a) temperature, (b) oxygen concentration, (c) solid flux and (d) density across the slurry with layer thicknesses between $150$ and $400 \ \mathrm{km}$. The dashed line in (a) is the uniform composition liquidus determined from \textit{ab initio} calculations \cite{Pozzo13,Davies15b}, and the dash-dotted line in (d) is the PREM density \cite{Dziewonski81}. The blue dotted line is the equivalent box model case from W18 for $d = 150 \ \mathrm{km}$ (see text). Control parameters are $Q^s = 2.5 \ \mathrm{TW}$, $Q^{sl}= 5.0 \ \mathrm{TW}$, $k = 100 \ \mathrm{W} \mathrm{m}^{-1} \mathrm{K}^{-1}$ and $T^{sl} = 5,547 \ \mathrm{K}$ (colour online).}
		\label{fig:profiles}
	\end{figure}
	
	Figure \ref{fig:profiles} shows example solutions for a wide range of layer thicknesses between $150$ and $400 \ \mathrm{km}$, where $Q^s = 2.5 \ \mathrm{TW}$ ($v_f = 0.44 \ \mathrm{mm} \mathrm{yr}^{-1}$), $Q^{sl}= 5 \ \mathrm{TW}$, $k = 100 \ \mathrm{W} \mathrm{m}^{-1} \mathrm{K}^{-1}$ and $T^{sl} = 5,547 \ \mathrm{K}$ taken from \cite{Pozzo13,Davies15b}, for an 82\%Fe-8\%O-10\%Si mixture. For all layer thicknesses the temperature gradient is strictly negative so the slurry is thermally destabilising, whereas the compositional gradient may either be stable or unstable. Solid flux is always negative in the direction towards the ICB, achieving its largest magnitude at the ICB itself and then vanishing to zero at the top of the layer as imposed by the boundary condition (\ref{eqn:nondim_jCSB}). Temperature, oxygen and solid flux all contribute to the overall density across the layer and the layer is stably-stratified when $150 \leq d \leq 300 \ \mathrm{km}$, however for $d \geq 350 \ \mathrm{km}$ the layer is unstable.
	
	We observe in Figure \ref{fig:profiles}(c) that the solid flux eventually increases exponentially with radius under the influence of turbulent mixing from the bulk of the liquid outer core. As the solid flux diminishes its gradient sharply increases, equivalent to precipitating more iron particles. The latent heat release associated with the phase change in the sublayer prompts the slurry to depress the temperature in Figure \ref{fig:profiles}(a) to stay on the liquidus and the oxygen concentration gradient increases in response to this as seen in Figure \ref{fig:profiles}(b).
	
	For completeness, we compare the solution in spherical geometry for $d = 150 \ \mathrm{km}$ with the equivalent boundary conditions from the Cartesian geometry of W18, except for the box model we have shifted the CSB temperature to coincide with the liquidus value from \cite{Pozzo13,Davies15b}. The main difference is that the effect of spherical geometry suppresses the effect of the mixing sublayer, since Figure \ref{fig:profiles}(c) shows that the solid flux gradient in the upper part of the layer is shallower than the box model case. Melting at the base of the layer in the box model is highlighted by the negative gradient in the solid flux, which is not present in the spherical model, and contributes positively to the density variations at the base of the layer in the box model. Despite this, turbulent mixing in the sublayer dominates to depress the temperature and oxygen variations, producing a much smaller density difference overall in the box model.
	
\begin{figure}[t]
		\centering
		\includegraphics[width=0.8\textwidth]{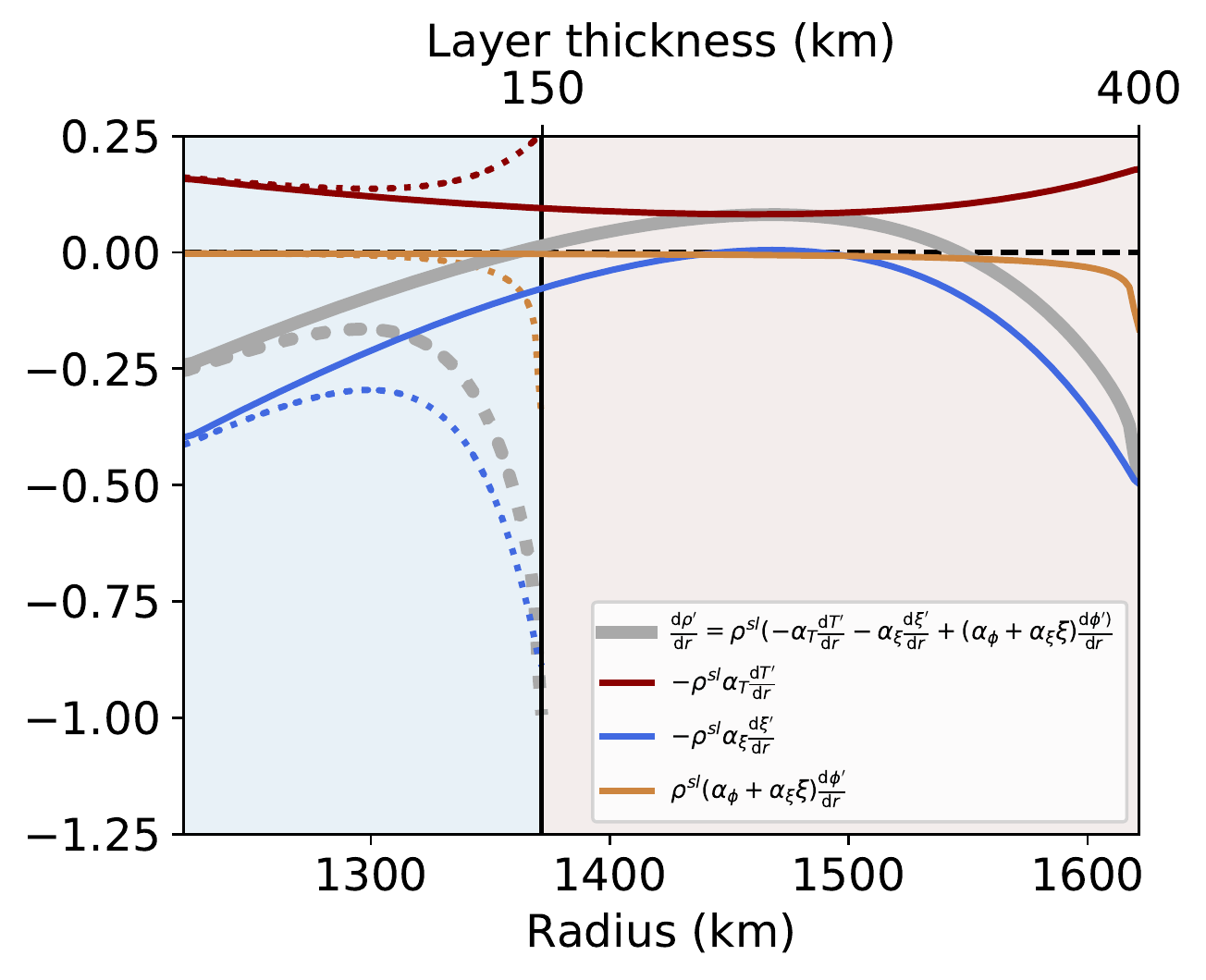}
		\caption{Gradients of the density variation (grey), $\mathrm{d}\rho'/\mathrm{d}r$, separated into contributions from the temperature (red), $-\rho^{sl}\alpha_T \mathrm{d}T'/\mathrm{d}r$, oxygen (blue), $-\rho^{sl}\alpha_\xi \mathrm{d}\xi'/\mathrm{d}r$, and solid fraction (orange), $\rho^{sl}(\alpha_\phi + \alpha_\xi \xi) \mathrm{d}\phi'/\mathrm{d}r$, across the slurry for layer thicknesses of $150$ (dotted) and $400 \ \mathrm{km}$ (solid). If $\mathrm{d}\rho'/\mathrm{d}r<0$ the slurry is stable, whereas if $\mathrm{d}\rho'/\mathrm{d}r>0$ the slurry is unstable. Control parameters are $Q^s = 2.5 \ \mathrm{TW}$, $Q^{sl}= 5.0 \ \mathrm{TW}$ and $k = 100 \ \mathrm{W} \mathrm{m}^{-1} \mathrm{K}^{-1}$ (colour online).} 
		\label{fig:eos}
	\end{figure}

	The control parameters $Pe$ and $Li_p$ depend linearly on $d$ and parameters $R_\rho$, $R_v$ and $Le$ depend indirectly on $d$ through $\rho^{sl}$. For a sense of scale in the variation of parameters from $d=150 \ \mathrm{km}$ to $400 \ \mathrm{km}$ for the reference solution given in Figure \ref{fig:profiles}, $Pe$ increases by $18\%$, $Li_p$ increases by $35\%$, $R_v$ increases by 14\% and the changes in $R_\rho$ and $Le$ are less than $1.5\%$. The combined effect of these changes strengthens the decay in temperature as a function of radius over the bulk of the layer (excluding the mixing sublayer), as seen in Figure \ref{fig:profiles}(a), but decreases $F$ and the strength of the turbulent mixing layer through the barodiffusion term in (\ref{eqn:nXi}), as seen in Figure \ref{fig:profiles}(c), which combine to reduce the magnitude of the density variations in the layer overall. Figure \ref{fig:eos} shows more clearly that contributions to $d\rho '/\mathrm{d}r$ from the oxygen gradient for $d = 400 \ \mathrm{km}$ can become positive in the mid-depths, since the pressure part of the liquidus relation (\ref{eqn:nLiq}) outweighs the temperature part, before becoming negative again under the stabilising influence of the mixing sublayer. Oxygen variations predominantly control the gradient of the density variations. When $d = 400 \ \mathrm{km}$ the sign of $\mathrm{d} \rho'/\mathrm{d}r$ changes from negative (stable), to positive (unstable), and back to negative again, producing an ``S"-shaped density profile seen in Figure \ref{fig:profiles}(d). On the other hand, Figure \ref{fig:eos} shows that for $d = 150 \ \mathrm{km}$ the layer is stable since $\mathrm{d} \rho'/\mathrm{d}r <0$ throughout the layer. Overall variations in the solid fraction, $\phi'$, are negligible over the majority of the layer, perhaps apart from a very thin region at the top (see \ref{sec:dphi} for further discussion on $\mathrm{d} \phi$).
	
	\subsection{Sensitivity analysis} \label{sec:sen}
\begin{figure}[t]
		\centering
		\includegraphics[width=0.8\textwidth]{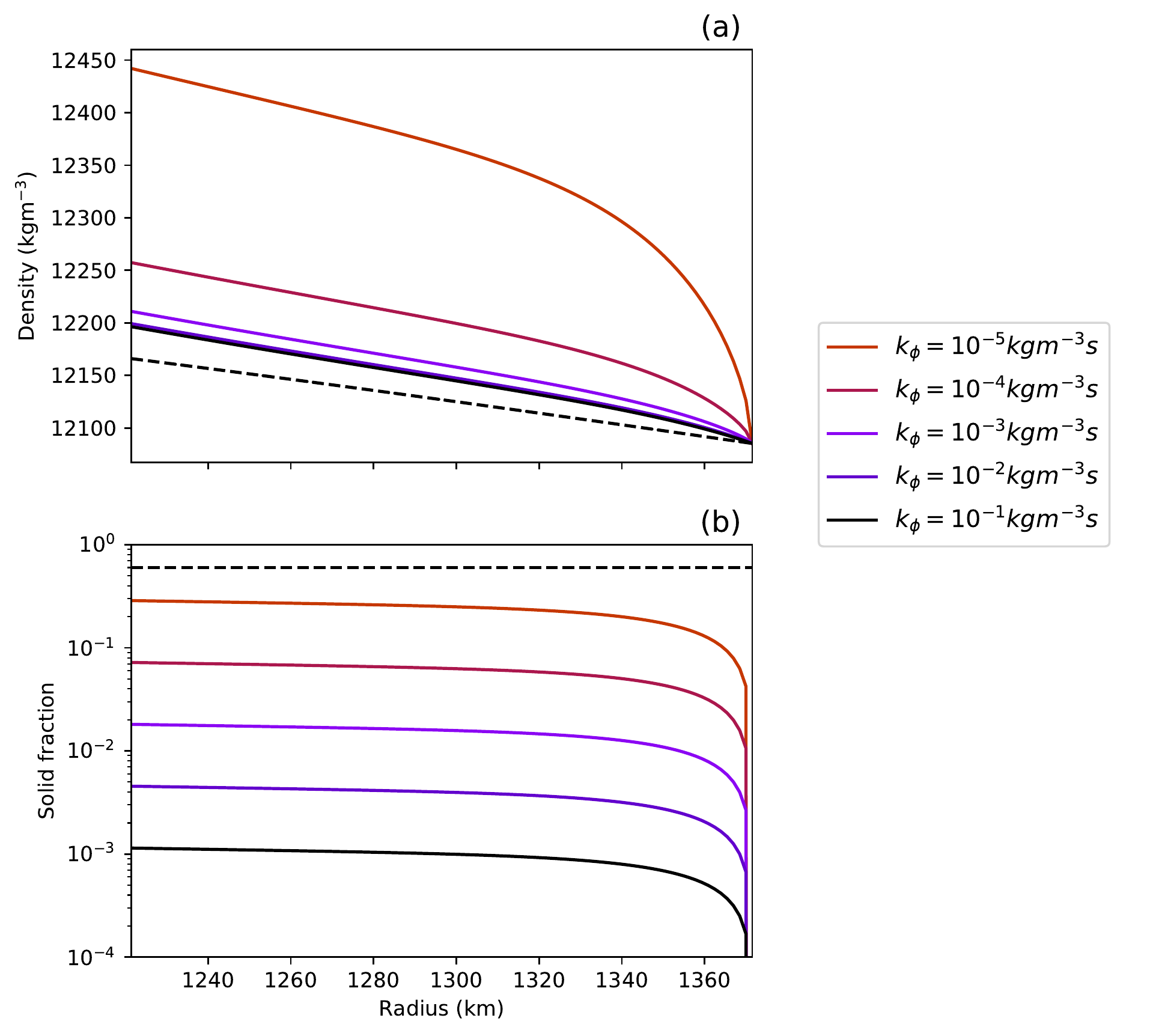}
		\caption{(a) Density and (b) solid fraction profiles across the layer for different values of $k_\phi$. The dashed line in (a) is PREM and the dashed in (b) is $\phi_m = 0.6$ that defines the rheological transition \cite{Solomatov15}. All other input parameters are the same as the case in Figure \ref{fig:profiles} with the layer thickness fixed at $150 \ \mathrm{km}$ (colour online).}
		\label{fig:sens_dens}
	\end{figure}

	We evaluate the sensitivity of the slurry system to the sedimentation prefactor, $k_\phi$, CSB temperature, $T^{sl}$, and CSB oxygen concentration, $\xi^{sl}$.	We vary the value of $k_\phi$ by a large range between $10^{-5}$ and $10^{-1} \ \mathrm{kg} \mathrm{m}^{-3} \mathrm{s}$ and recompute the example given in Figure \ref{fig:profiles}(a) where the layer thickness is fixed at $150 \ \mathrm{km}$. Figure \ref{fig:sens_dens}(a) shows that density stratification is increased with smaller values of $k_\phi$ and the solutions converge as $k_\phi$ exceeds $10^{-3} \ \mathrm{kg} \mathrm{m}^{-3} \mathrm{s}$. Figure \ref{fig:sens_dens}(b) shows the corresponding solid fraction, $\phi$, profiles within the layer. It can be seen that for the smallest value of $k_\phi$, the solid fraction is close to the rheological transition at $\phi_m = 0.6$ \cite{Solomatov15}, which violates the assumption that $\phi \ll 1$ in the slurry. We select a value of $k_\phi = 10^{-2} \ \mathrm{kg} \mathrm{m}^{-3} \mathrm{s}$ that is high enough so that the density becomes independent of $k_\phi$ and consistent with $\phi \ll 1$ in the slurry.
	
	For the CSB temperature, we recompute the example case with $d = 150 \ \mathrm{km}$ and vary $T^{sl}$ between $4,500$ and $6,000 \ \mathrm{K}$ with intervals of $100 \ \mathrm{K}$ and for the CSB oxygen concentration we vary its value between $2.0$ and $12.0 \ \textrm{mol.}\%$ with intervals of $0.5 \ \textrm{mol.}\%$.
	\begin{figure}[t]
		\centering
		\includegraphics[width=\textwidth]{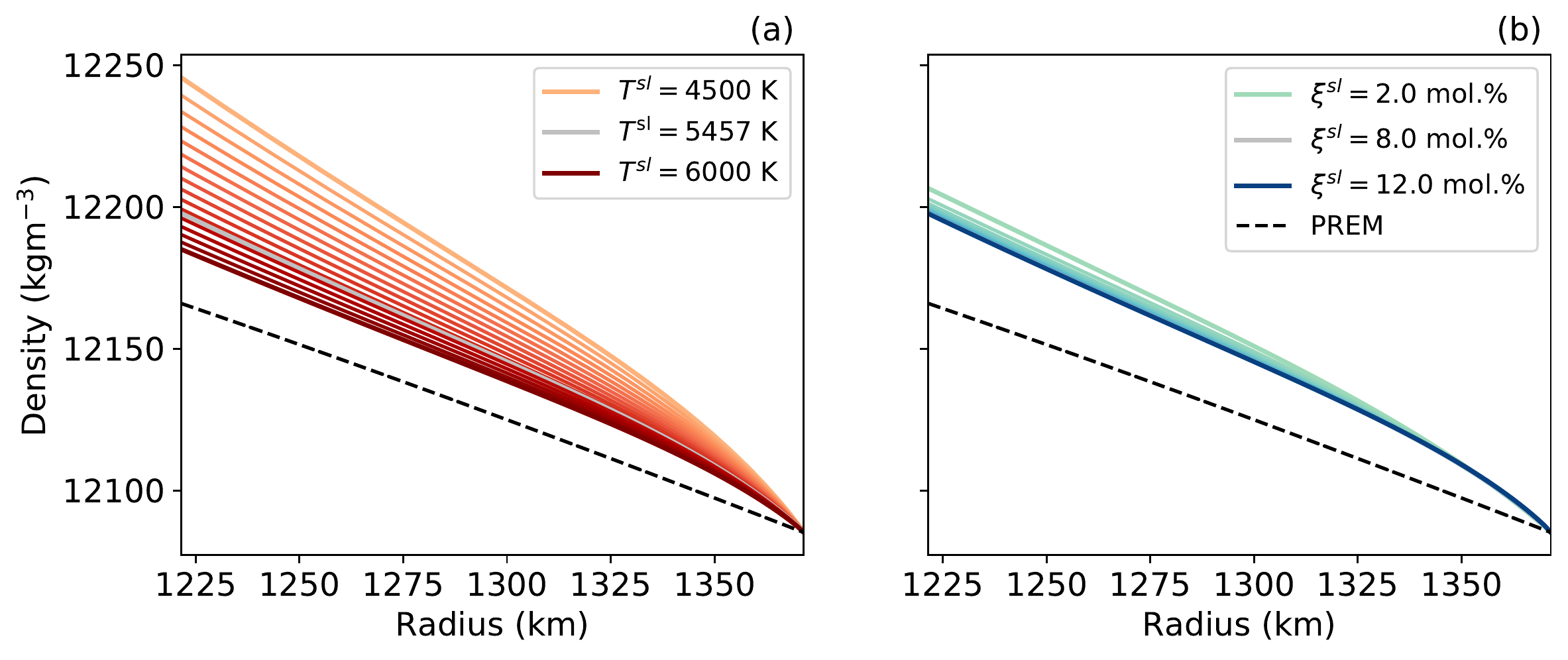}
		\caption{Density profiles across the layer for (a) $4,500 < T^{sl} < 6,000 \ \mathrm{K}$ and (b) $2.0 < \xi^{sl} < 12.0 \ \mathrm{mol.}\%$. Default values of $T^{sl} = 5,547 \ \mathrm{K}$ from \cite{Pozzo13,Davies15b} in (a) and $\xi^{sl} = 8.0 \ \mathrm{mol.}\%$ in (b) are given by the grey lines. All other input parameters are the same as the example case, and the layer thickness is fixed at $150 \ \mathrm{km}$ (colour online).}
		\label{fig:sens}
	\end{figure}
	Figure \ref{fig:sens}(a) and (b) presents the density profiles across the layer with varying $T^{sl}$ and $\xi^{sl}$, respectively. We find that no solutions were obtained for values below $\xi^{sl} = 2.0 \ \textrm{mol.}\%$. Figure \ref{fig:sens}(b) shows that despite the variation in $\xi^{sl}$, there appears to be a limited effect on the resulting density profiles, whereas Figure \ref{fig:sens}(a) shows that changing $T^{sl}$ introduces a greater spread in the solutions obtained. In terms of the dimensionless parameters, varying $2.0 \ \mathrm{mol.}\% < \xi^{sl} < 12.0 \ \mathrm{mol.}\%$ through $0.004 < Li_\xi < 0.028$ does not significantly affect the system, where $Li_\xi$ enters into the pressure part of the liquidus relation and the barodiffusion term of equation (\ref{eqn:nXi}). Changing $T^{sl}$ in boundary condition (\ref{eqn:nondim_tempCSB}) shifts the anchor point of the temperature solution and will not affect the curvature of the temperature profile since the heat fluxes into and out of the slurry remain fixed, so the temperature perturbations are the same for all $T^{sl}$. However, $T^{sl}$ does affect the oxygen concentration through the liquidus relation, where a higher $T^{sl}$ will decrease the oxygen concentration gradient and create smaller perturbations in the oxygen concentration, therefore generating smaller density perturbations to produce a lower slurry density overall. A lower $T^{sl}$ has the opposite effect to produce a higher slurry density. The spread in the ICB density with different $T^{sl}$ is roughly $60 \ \mathrm{kg} \mathrm{m}^{-3}$, and there is up to a 40\% change in the density jump across the whole layer with respect to the example case. This sensitivity may have an impact on the overall stability of the layer, however for the regime diagram we shall continue to use the default value from the literature of $T^{sl} = 5,547 \ \mathrm{K}$ for a layer $150 \ \mathrm{km}$ thick.
	
	\subsection{Regime diagram} \label{sec:reg}
	We present a regime diagram of the $Pe,St$--space by varying $Q^s$ and $Q^{sl}$. There are two cases: high thermal conductivity, $k = 100 \ \mathrm{W} \mathrm{m}^{-1} \mathrm{K}^{-1}$, corresponding to $Le = 1181$ and low thermal conductivity, $k = 30 \ \mathrm{W} \mathrm{m}^{-1} \mathrm{K}^{-1}$, corresponding to $Le = 354$. By fixing $d = 150 \ \mathrm{km}$ and $\xi^{sl} = 8 \ \mathrm{mol.}\%$, then $Li_p = 0.16$, $Li_\xi = 0.018$, $R_\rho = 0.952$ and $R_v = 0.235$.
	
	\begin{figure}[t]
		\centering
		\includegraphics[width=\textwidth]{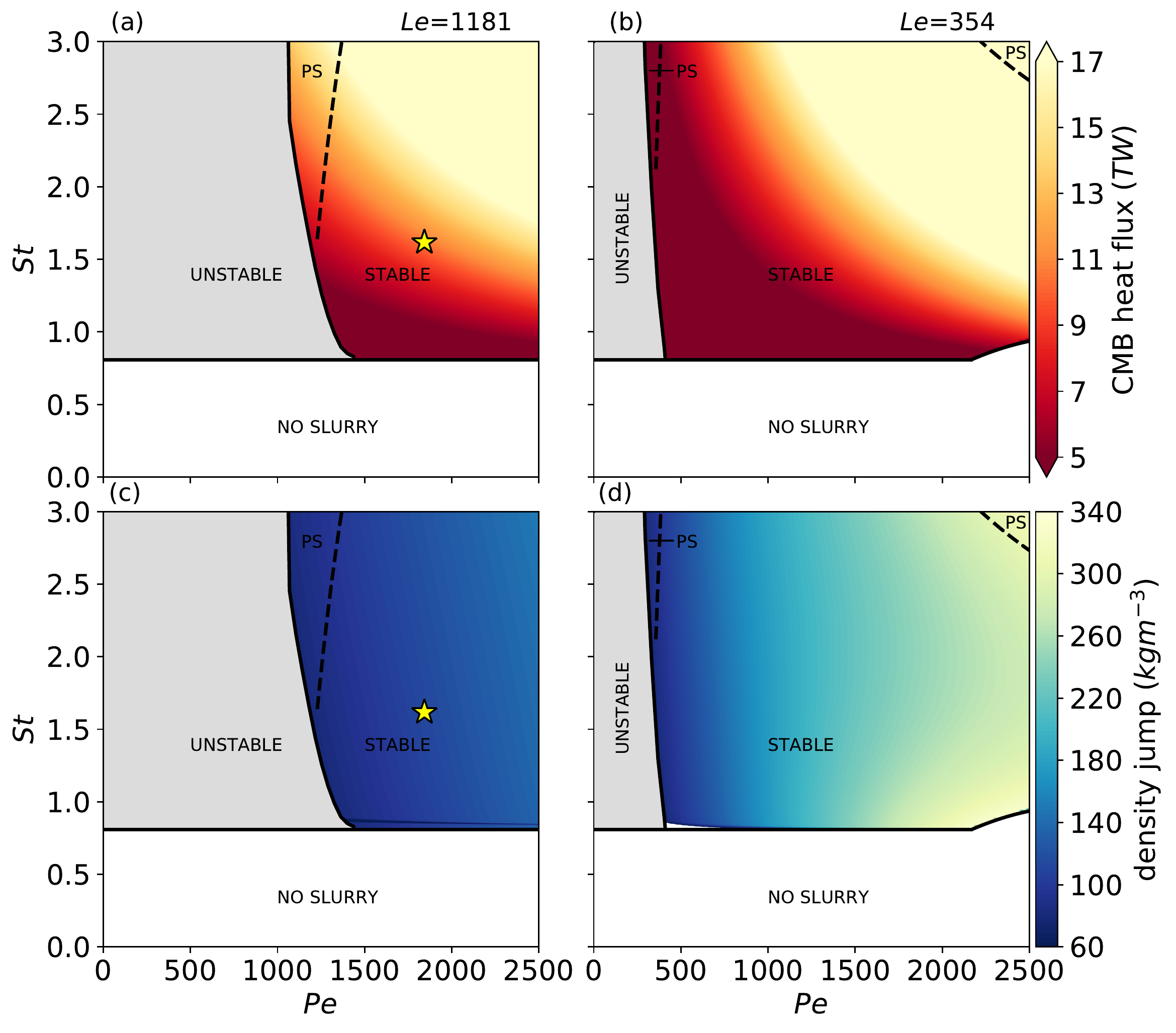}
		\caption{(Left column) Regime diagrams for the high Lewis number case, (right column) and the low Lewis number case with $d = 150 \ \mathrm{km}$. (Top row) Contours of the CMB heat flux constrained by $5 < Q^c < 17 \ \mathrm{TW}$ and (bottom row) contours of the density jump, $\rho^s_+ - \rho^{sl}$. The phase space is divided into stable (contour fill), partially stable, denoted PS (contour fill, dashed line), unstable (grey) and no slurry (white) regions. The yellow star represents the example case from Figure \ref{fig:profiles} (colour online).}
		\label{fig:regime}
	\end{figure}
	
	Figure \ref{fig:regime}(a) and (c) presents the regime diagram for the high Lewis number case, and Figure \ref{fig:regime}(b) and (d) is the regime diagram for the low Lewis number case. Regions of the phase space are divided into stable ($\mathrm{d} \rho' / \mathrm{d} r <0$), partially stable ($\mathrm{d} \rho' / \mathrm{d} r <0$ for at least $100 \ \mathrm{km}$) and unstable ($\mathrm{d} \rho' /\mathrm{d} r >0$) slurries, and also areas with no slurry where no solution is found.
	
	No slurry can exist when the latent heat release at the ICB interface exceeds the CSB heat flux, so that
	\begin{align*}
		\textrm{if} \qquad Q^{sl} < Q^s \qquad \textrm{then} \qquad
		St = \frac{q^{sl}}{q^s} < \left(\frac{r^i}{r^{sl}}\right)^2 \equiv St^*
	\end{align*}
	which corresponds to a critical Stefan number of $0.6 \leq St^* \leq 0.8$ for $150 \ \mathrm{km} \leq d \leq 400 \ \mathrm{km}$. When the layer thickness is $150 \ \mathrm{km}$, above $St^* = 0.8$ we find a region of unstable solutions at lower $Pe$ and stable solutions at higher $Pe$, and this transition generally occurs when
		\begin{align}
		Pe_T \equiv \frac{Pe}{Le} = \frac{v_f r^{sl}}{\kappa} \simeq 1, \label{eqn:PeT}
		\end{align}
	where $Pe_T$ is the thermal P\'{e}clet number and $\kappa = k / \rho ^{sl} c_p$ is the thermal diffusivity. $Pe_T$ controls the boundary condition on the temperature gradient in (\ref{eqn:nondim_tempgradICB}) and (\ref{eqn:nondim_tempgradCSB}). An unstable slurry develops when $Pe < Le$ because more heat is conducted through the slurry which is unavailable for equilibration through the liquidus constraint, hence creating a smaller oxygen concentration difference that produces a smaller density contrast that is insufficient to stabilise the layer. A stable slurry develops when $Pe > Le$ as the rate of advection from the compacting layer overcomes the thermal diffusion rate. Increasing $Pe_T$ stabilises the slurry layer since the temperature gradient steepens at the boundaries, therefore generating greater positive variations in the oxygen gradient that enhance the density anomaly. If $d=150 \ \mathrm{km}$ then the transition at $Pe \simeq Le$ corresponds to $Q^s = 1.5 \ \mathrm{TW}$ for high $Le$ and $Q^s = 0.5 \ \mathrm{TW}$ for low $Le$. Figure \ref{fig:regime}(c) and (d) show that the density jump across the layer in all cases is well below the maximum limit of $720 \ \mathrm{kg} \mathrm{m}^{-3}$ and also suggests that the density jump roughly scales with $Pe$, with higher values encountered at low $Le$.
	
	A higher $St$ steepens the CSB temperature gradient relative to the ICB, and condition (\ref{eqn:PeT}) is met by strengthening the turbulent sublayer at the top of the slurry by increasing the mixing parameter $F$. Within the sublayer the rate of crystallisation is increased, therefore the density jump across the layer increases with $St$ as seen in Figure \ref{fig:regime}(c) and (d). The critical $Pe_T$ where the slurry transitions from unstable to stable slightly decreases as $St$ increases and the layer is more likely to become partially stable. We find that the dimensionless ICB speed, and thereby the magnitude of the solid flux at the ICB, is roughly proportional to $St$. The ICB speed affects contributions to the secular cooling, gravitational power and latent heat release in the core and therefore influences the total CMB heat flow. This is reflected in Figure \ref{fig:regime}(a) and (b) on the total CMB heat flux, which scales linearly with $PeSt$.
	
	\subsection{Seismic properties of the F-layer and the inner core age}
	\begin{figure}[H]
		\centering
		\includegraphics[width=0.8\textwidth]{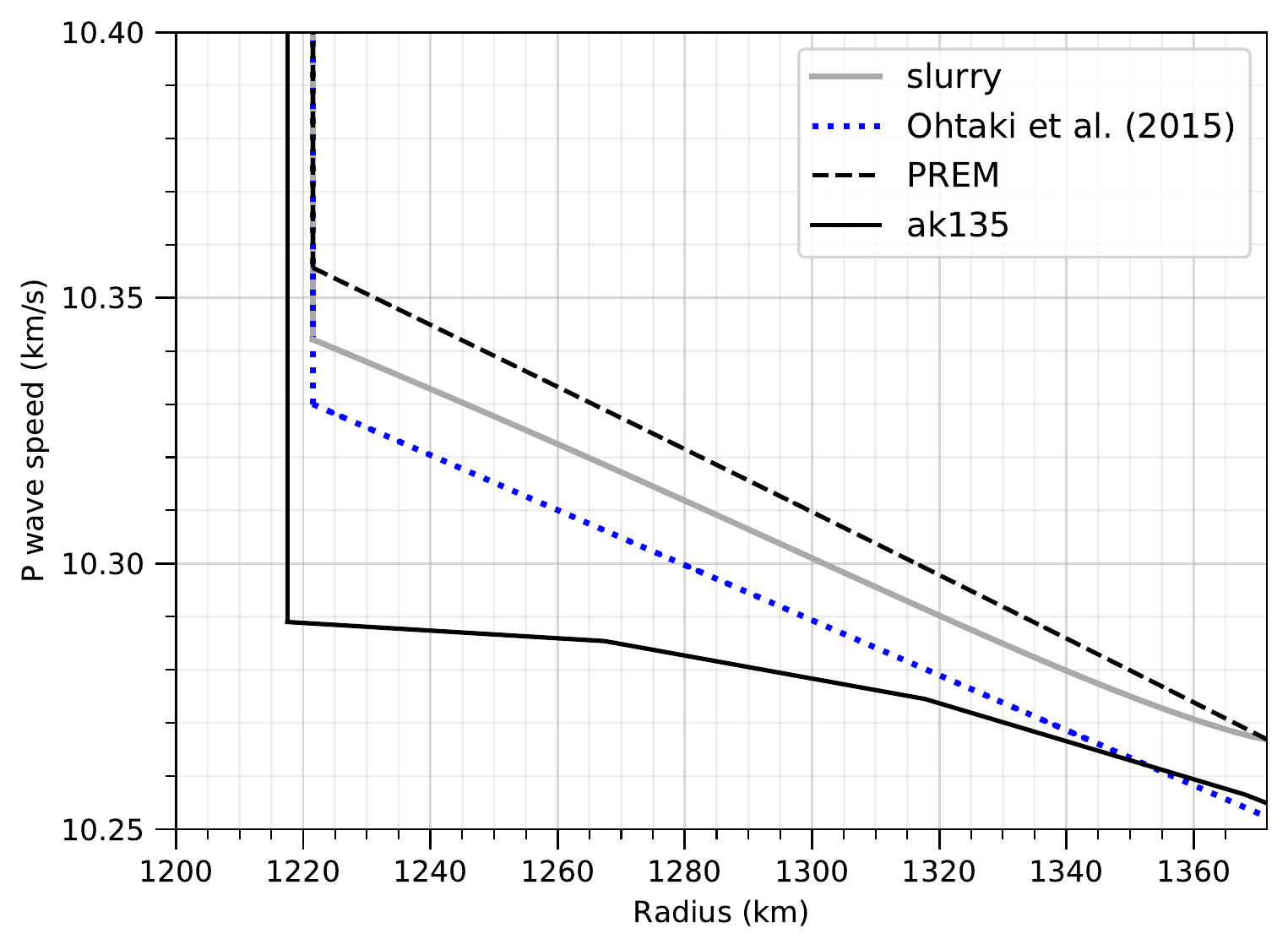}
		\caption{P wave speed in the F-layer derived from the high $Le$ example solution with $d = 150 \ \mathrm{km}, \ k = 100 \ \mathrm{W} \mathrm{m}^{-1} \mathrm{K}^{-1}, \ Q^s = 2.5 \ \mathrm{TW}$ and $Q^{sl} = 5.0 \ \mathrm{TW}$ (grey, solid). Seismic models are PREM (black, dashed), ak135 (black, solid) and the FVW model of Ohtaki \textit{et al.} \cite{Ohtaki15} (blue, dotted) (colour online).}
		\label{fig:seismic}
	\end{figure}
	Motivated by the seismic observations, we have demonstrated that a slurry is able to produce stable layers with varying degrees of stratification depending on the parameters selected. If we invert the process, then what constraints can the model provide on observations? From our slurry model we can determine the density at every point across the layer and hence the P wave speed, $v_p^2 = K/\rho$. The bulk modulus, $K$, depends on the core chemistry and therefore an appropriate equation of state should be applied \cite{Badro14,Cottaar14}. For the sake of simplicity we approximate the bulk modulus by PREM in our calculations.
	
	Figure \ref{fig:seismic} illustrates the $v_p$ profile of the example solution from Section \ref{sec:example} compared with the seismic models of PREM \cite{Dziewonski81}, ak135 \cite{Kennett95} and the FVW model of Ohtaki \textit{et al.} \cite{Ohtaki15}. The P wave speed from the slurry model is reduced relative to PREM by up to $0.13 \%$ and there is a difference of $0.1 \%$ compared with Ohtaki \textit{et al.} \cite{Ohtaki15}, which reported that $v_p(r^i) = 10.3 \ \mathrm{km} \ \mathrm{s}^{-1}$ and $\mathrm{d} v_p / \mathrm{d} r = -5.2 \times 10^{-7} \ \mathrm{s}^{-1} $. For this example slurry solution we obtain a density jump across the layer of $\rho^s_+ - \rho^{sl} = 112 \ \mathrm{kg} \mathrm{m}^{-3}$ and calculate that $Q^c = 10.4 \ \mathrm{TW}$, which is within the acceptable range of CMB heat flows. On the slurry side of the ICB we have $\rho^s_+ = 12,198 \ \mathrm{kg} \mathrm{m}^{-3}$, therefore we predict that the density jump from body waves for this particular slurry is $\Delta \rho_\textrm{bod} = \rho^s_- - \rho^s_+  = 502 \ \mathrm{kg} \mathrm{m}^{-3}$. The density jump from normal modes is $\Delta \rho_\textrm{mod} = \Delta \rho_\textrm{bod} + \rho^s_+ - \rho^{sl} = 628 \ \mathrm{kg} \mathrm{m}^{-3}$, and this is fixed in all solutions of the slurry by choosing $\rho_-^s$ and $\rho^{sl}$ from PREM.
	
	Note that the example solution given above is one of many compatible solutions given by the regime diagram. By considering the entire range of parameters permitted by the regime diagram that fit with the geophysical constraints, the high $Le$ case gives $81 \leq \rho^s_+ - \rho^{sl} \leq 140 \ \mathrm{kg} \ \mathrm{m}^{-3}$ and the low $Le$ case gives $101 \leq \rho^s_+ - \rho^{sl} \leq 331 \ \mathrm{kg} \ \mathrm{m}^{-3}$. This leads to bounds of $475 \leq \Delta \rho_\textrm{bod} \leq 534 \ \mathrm{kg} \ \mathrm{m}^{-3}$ and $283 \leq \Delta \rho_\textrm{bod} \leq 513 \ \mathrm{kg} \ \mathrm{m}^{-3}$ for high and low $Le$ respectively. The observations available give $520 \pm 240 \leq \Delta \rho_\textrm{bod} \leq 1,100 \ \mathrm{kg} \ \mathrm{m}^{-3}$, therefore our model suggests that a stably-stratified F-layer is opposed to the higher values of $\Delta \rho_\textrm{bod}$ recommended by the observations and that the slurry model gives an upper bound of $\Delta \rho_\textrm{bod} \leq 534 \ \mathrm{kg} \ \mathrm{m}^{-3}$ for both $Le$ cases.
	
	We can approximate the inner core age using the model output of the ICB speed, $v$, by assuming that inner core growth is proportional to the square-root of time \cite{Labrosse14},
	\begin{align}
		r^i(\tau) = r^i_0 \left(1+\frac{\tau}{\tau^i}\right)^\frac{1}{2}, \label{eqn:ic_growth}
	\end{align}
	where $\tau$ is time normalised by the age of the inner core, $\tau^i$, and $r^i_0$ is the present-day IC radius. By differentiating (\ref{eqn:ic_growth}) and considering the present-day where $\tau=0$, we have
	\begin{align}
		\tau^i = \frac{r^i_0}{2 v}, \label{eqn:ic_age}
	\end{align}
	where we recall that $v \equiv v_f + v_s = \mathrm{d} r^i/\mathrm{d}\tau$ is the inner core growth rate. Estimating $\tau^i$ is of significant geophysical interest and its value is subject to debate since $\tau^i$ is directly related to the evolution of the core global heat balance, which is significantly affected by the thermal conductivity \cite{Williams18}. For the same range of solutions considered above to estimate $\Delta \rho_\mathrm{bod}$, our results from the regime diagram for high $Le$ give $0.52  \leq v \leq 0.98 \ \mathrm{mm} \ \mathrm{yr}^{-1}$, which is relatively fast compared to the low $Le$ solutions where $0.24  \leq v \leq 0.31 \ \mathrm{mm} \ \mathrm{yr}^{-1}$. This yields an inner core age of $0.6 \leq \tau^i \leq 1.2 \ \mathrm{Ga}$ for high $Le$, which is in line with other estimates from the literature \cite{Nimmo15a}, and $2.0 \leq \tau^i \leq 2.6 \ \mathrm{Ga}$ for low $Le$, which is mostly beyond oldest estimates of $2.0 \ \mathrm{Ga}$ from the literature using low thermal conductivity \cite{Davies15b}. Our results therefore suggest a slurry model with high core conductivity predicts geophysical properties of the F-layer and core that are consistent with independent seismic and geodynamic calculations.
	
	\section{Conclusions} \label{sec:con}
	In this study, we have investigated the conditions that produce a stably-stratified slurry layer at the base of the Earth's outer core. We non-dimensionalise the governing equations given by W18 and elucidate the key dimensionless parameters that control the system behaviour. By varying the P\'{e}clet and Stefan numbers, we map a regime diagram of the slurry demarcating the conditions that favour stable, partially stable  and unstable density configurations, as well as no slurry. Solutions are obtained for high and low $Le$ numbers, which reflect high or low thermal conductivity values of the core. We constrain the results by evaluating the density jump across the layer and the total CMB heat flux. 
	
	Our main result is that stably-stratified solutions can be produced by a slurry for a wide range of parameters that span plausible values for Earth's core. We have identified regions of the parameter space containing many solutions that are compatible with observations of the F-layer, and we have also established conditions that are not suitable. We find that
	\begin{itemize}
		\item $Pe_T = \frac{Pe}{Le} \simeq 1$ divides the space between unstable and stable slurry layers,
		\item higher P\'{e}clet number slurries generally facilitate stable density stratification because this controls the steepness of the temperature gradient at the boundaries, which in turn increases the magnitude of density variations in the slurry,
		\item the Stefan number has a stabilising role through crystallising more particles in the turbulent mixing sublayer,
		\item no slurry can exist with $St < St^* = (r^i/r^{sl})^2$ since the heat flow at the bottom of the layer is greater than at the top of the layer
		\item the slurry model suggests that $\Delta \rho_\textrm{bod} \leq 534 \ \mathrm{kg} \ \mathrm{m}^{-3}$ for high and low $Le$
		\item estimates of the inner core age suggests a slurry with high core conductivity is compatible with independent estimates from the literature
	\end{itemize}
	
	Density perturbations of compositional origin dominate contributions to the overall stability of the layer, and can be destabilised through increasing the layer thickness. We have also investigated the sensitivity to the CSB temperature and oxygen concentration and deduced a limit for the prefactor $k_\phi = 10^{-2} \ \mathrm{kg} \mathrm{m^{-3}} \mathrm{s}$ that defines the sedimentation coefficient, $b(\phi)$, which relates the solid flux, $\mathbf{j}$, to the solid fraction, $\phi$, through Stokes' flow. We find that the model is insensitive to the concentration of oxygen at the CSB, however, the effect of the liquidus temperature is significant to the overall stability of the layer.
	
	From mineral physics or seismology of model uncertainties, improved estimates such as the layer thickness, CSB temperature and the CSB oxygen concentration, will benefit the slurry model greatly. Further progress determining the liquidus curve for an iron mixture, using experiments or first-principle calculations, could improve the temperature condition at the CSB, and improved estimates of the density jump across the layer from normal and body wave studies would help constrain the space of geophysically consistent solutions.
	
	We examined the P wave speed in the slurry layer to show that the model can produce a profile that compares well with other seismic models. This could be further verified comprehensively so that the model may provide a useful tool for corroborating observations, such as constraining the seismic density jump at the ICB derived from body waves, $\Delta \rho_\textrm{bod}$. There is also the potential for the slurry to provide improved and independent estimates of the inner core age stipulated by having a stable F-layer that agrees with the seismic and heat flow requirements.
	
	We implemented a simple compacting layer on the solid side of the IC where solid particles produced by the slurry accumulate and instantly compact. Realistically the physics of this process is extremely complicated. Studies suggest that a mush solidification regime can occur \cite{Deguen12} where liquid channels permeate a matrix of solid with a high solid fraction. A more advanced model may seek to incorporate this process. The timescale of interest in our slurry is comparable to the slow growth of the inner core, where we have applied the fast melting limit. On short timescales at the microscopic level, the effect of supercooling on the nucleation of solid iron at core conditions may come into play and this topic is an active area of interest \cite{Huguet18,Davies19,Lasbleis19}.
	
	Future work could focus on assessing the dominant balances between terms in the slurry equations that are responsible for stable density stratification so that the physics controlling the transitions between different regimes can be elucidated. For example, the importance of the mixing sublayer and its influence on the production of solid phase needs to be quantitatively evaluated. A more sophisticated model coupling the slurry to outer core convection, perhaps similar to a recent study conducted by Bouffard \textit{et al.} \cite{Bouffard20} for the stably-stratified layer at the top of the core, may fully capture the physics of this process and shed light on the effect of entrainment.
	
	Our study also opted to keep the layer thickness constant over time in order to ascertain solutions that are concomitant with the present-day seismic observations. Relaxing the layer thickness requires an extra constraint on the model to be developed in its place \cite{Wong18b}. Constructing a fully time-dependent framework could provide insight into the factors controlling the growth and decline of a slurry layer, which may shed light on how the F-layer came into existence over the core's history. 
	
	\section{Acknowledgements}
	The numerical code used to solve the slurry equations in this paper is freely available at \url{https://github.com/jnywong/nondim-slurry}. We thank Marine Lasbleis and an anonymous reviewer for their constructive comments that helped improve this paper. JW acknowledges support from the Fondation Simone and Cino Del Duca of Institut de France and the Engineering and Physical Sciences Research Council (EPSRC) Centre for Doctoral Training in Fluid Dynamics (EP/L01615X/1). CJD is supported by the Natural Environment Research Council (NERC) Independent Research Fellowship (NE/L011328/1). Figures were produced using Matplotlib \cite{Hunter07}.
	
	\newpage

	\appendix
	\section{Table of values} \label{sup:val}
	
	\begin{longtable}{| p{.10\textwidth} | p{.30\textwidth} | p{.20\textwidth} | p{.15\textwidth} | p{.20\textwidth} |}
		\hline
		\textbf{Symbol} & \textbf{Definition} & \textbf{Value} & \textbf{Units} & \textbf{Source} \\
		\hline
		% --------Group 1a-----------------------------
		$p$     & Pressure     & & $\mathrm{kg} \mathrm{m}^{-1} s^{-2}$     & \\
		$T$     & Temperature  & & $\mathrm{K}$                             & \\
		$\xi$   & Light element concentration &  & Mass fraction    & \\
		$\xi^l$ & Light element concentration in the liquid phase  & & Mass fraction   & \\
		$\phi$  & Solid fraction & & Mass fraction                           & \\
		$j$     & Solid flux    & & $\mathrm{kg}\mathrm{m}^{-2}\mathrm{s}^{-1}$   & \\
		$\rho$  & Density       & & $\mathrm{kg}\mathrm{m}^{-3}$             & \\
		\hline
		% --------Group 1b-----------------------------
		$F$     & Mixing parameter &                            &           & \\
		$v$   & ICB speed  &  & $\mathrm{m}\mathrm{s}^{-1}$              & \\
		$v_f$   & Freezing speed  &  & $\mathrm{m}\mathrm{s}^{-1}$              & \\
		$v_s$   & Snow speed  &  & $\mathrm{m}\mathrm{s}^{-1}$              & \\
		$\mu$   & Chemical potential & & $\mathrm{J}\mathrm{kg}^{-1}$         & \\
		$\rho_H$  & Hydrostatic density &  & $\mathrm{kg} \mathrm{m^{-3}}$  & \\
		$\rho'$ & Density perturbation &  & $\mathrm{kg} \mathrm{m^{-3}}$   & \\
		$T'$    & Temperature perturbation &  & $\mathrm{K}$                & \\
		${\xi^l}'$  & Light element perturbation in the liquid phase &  & Mass fraction & \\
		$\phi'$ & Solid fraction perturbation &  & Mass fraction            & \\
		$b(\phi)$ & Sedimentation coefficient &  & $\mathrm{kg} \mathrm{m^{-3}} \mathrm{s}$ & \\
		$k_\phi$ & Sedimentation coefficient prefactor &  &$\mathrm{kg} \mathrm{m^{-3}} \mathrm{s}$ & \\
		$\nu$   & Kinematic viscosity &  & $\mathrm{m}^2 \mathrm{s}^{-1}$   & \\
		$N$     & No. of solid iron particles per unit volume &  &          & \\
		$\Phi$  & Gibbs free energy   &  &                                  & \\
		\hline
		% --------Group 1d-----------------------------
		$Pe$ & P\'{e}clet number & & &  \\
		$St$ & Stefan number & & &  \\
		$Le$ & Lewis number & & &  \\
		$Li_p$ & Liquidus number (pressure) & & &  \\
		$Li_\xi$ & Liquidus number (composition) & & &  \\
		$R_\rho$ & Ratio between liquid and solid iron & & &  \\
		$R_v$ & Ratio between the change in specific volumes upon phase change from pure iron and iron alloy & & &  \\
		
		\hline
		% --------Group 2a-----------------------------
		$R$     & Ideal gas constant & $8.31$ & $\mathrm{J} \mathrm{kg}^{-1} \mathrm{mol.}^{-1}$ &  \\
		$a_O$ & Atomic weight of oxygen & $16$ & $\mathrm{Da}$ &  \\
		\hline
		% --------Group 2b-----------------------------
		$r^{i}$ & ICB radius & $1221 \times 10^3$ & $\mathrm{m}$& PREM \citep{Dziewonski81} \\
		$d$ & Layer thickness & $(150,200,250$, $300,350,400)$ $\ \times 10^3$ & $\mathrm{m}$ & \cite{Souriau91}, \cite{Zou08} \\
		$r^{sl}$ & CSB radius & $1371$--$1621 \ \times 10^3$ & $\mathrm{m}$& \cite{Souriau91}, \cite{Zou08} \\
		% $\rho_P$ & PREM density &  & $\mathrm{kg} \mathrm{m}^{-3}$ & PREM \citep{Dziewonski81}  \\
		$\rho^s_-$  & Density of iron on the solid side of the ICB & $12.76 \times 10^3$  & $\mathrm{kg}\mathrm{m^{-3}}$ & PREM \citep{Dziewonski81} \\
		$\rho^s_+$  & Density of iron on the liquid side of the ICB &  & $\mathrm{kg} \mathrm{m^{-3}}$ & \\
		$\rho ^{sl}$ & Density of liquid iron at $r = r^{sl}$ & & $\mathrm{kg} \ \mathrm{m^{-3}}$ & PREM \citep{Dziewonski81} \\
		$g$      & Gravity &  & $\mathrm{m} \mathrm{s}^{-2}$ & PREM \citep{Dziewonski81}  \\
		$K$    & Bulk modulus &  & $\mathrm{kg} \mathrm{m}^{-1} s^{-2}$ & PREM \citep{Dziewonski81}  \\
		\hline
		% --------Group 3-----------------------------
		$\xi^{sl}$ & Oxygen concentration in the bulk of the liquid core & $2$--$12$ & $\mathrm{mol.}\%$ & \cite{Hirose13} \\
		$T^{sl}$   & Liquidus temperature at the CSB & $4,500$--$6,000$ & $\mathrm{K}$ & \cite{Davies15b, Pozzo13} \\
		$c_p$ &Specific heat capacity & $715$ &$\mathrm{J} \mathrm{kg}^{-1}\, \mathrm{K}^{-1}    $&\cite{Gubbins03}\\
		$\alpha$ & Thermal expansion coefficient & $1 \times 10^{-5}$ & $\mathrm{K}^{-1}$ & \cite{Gubbins03} \\
		$\alpha_\xi$ & Compositional expansion coefficient of oxygen & $1.1$ && \cite{Gubbins04b} \\
		$L$ & Latent heat of fusion & $0.75 \times 10^6$ & $\mathrm{J} \mathrm{kg}^{-1}$ &         \cite{Gubbins03}\\
		$k$ & Thermal conductivity & $30$, $100$ & $\mathrm{W} \mathrm{m}^{-1} \mathrm{K}^{-1}$ & \cite{Williams18}, \cite{Davies15b} \\
		$\bar{D}$ & Modified self--diffusion coefficient of oxygen &  &$\mathrm{m}^2 \mathrm{s}^{-1}$ & \\
		$D_O$ & Self--diffusion coefficient of oxygen & $0.98 \times 10^{-8}$ &$\mathrm{m}^2 \mathrm{s}^{-1}$ & \cite{Pozzo13}\\
		
		\hline
		% --------Group 4-----------------------------
		%$\rho\lO$ & Specific density of light element & $5.56 \times 10^3$ & $\mathrm{kg}                         \mathrm{m^{-3}}$ & \cite{Gubbins04b} \\
		% $\rho^{sl}$ & Specific density of liquid iron, reference density at CSB &  & $                    \mathrm{kg} \mathrm{m^{-3}}$ &  PREM \cite{Dziewonski81} \\
		$V\sFe$ & Specific volume of solid iron &  & $\mathrm{m^{3}} \mathrm{kg}^{-1} $ & Ideal solution theory \\
		$V\lFeO$ & Specific volume of liquid iron and oxygen &  & $\mathrm{m^{3}} \mathrm{kg}^{-1}$ & Ideal solution theory  \\
		%$\Delta V^{sl}O$ & Change in specific volume between light element and liquid iron & & $\mathrm{kg}^{-1} \mathrm{m}^3$&  Ideal solution theory \\
		$\Delta V\slFeO$ & Change in specific volume between liquid and solid phase &                 &$\mathrm{m}^3 \mathrm{kg}^{-1} $ & Ideal solution theory \\
		$\Delta V\slFe$  & Change in specific volume between liquid iron and solid iron & & $\mathrm{m}^3 \mathrm{kg}^{-1} $  &  Ideal solution theory \\
		$\alpha_\phi$ & Expansion coefficient of solid & & & Ideal solution theory  \\
		
		\hline
		% --------Group 5-----------------------------
		%    $v_f$ & Freezing speed &  & $\mathrm{m} \mathrm{s}^{-1} $ & \\
		$ q^s$ & ICB heat flow per unit area & & $\mathrm{W}\mathrm{m}^{-2}$ &\\
		$ q^{sl}$ & CSB heat flow per unit area & & $\mathrm{W}\mathrm{m}^{-2}$ &\\
		$ Q^s$ & ICB heat flow & & $\mathrm{TW}$ &\\
		$ Q^{sl}$ & CSB heat flow & & $\mathrm{TW}$ & \\
		$ Q^c$ & CMB heat flow & $5$--$17$ & $\mathrm{TW}$ & \cite{Lay08}, \cite{Nimmo15a} \\
		$ Q_{s}$ & Secular cooling & & $\mathrm{TW}$ & \\
		$ Q_{g}$ & Gravitational power & & $\mathrm{TW}$ & \\
		$ Q_{l}$ & Latent heat power & & $\mathrm{TW}$ & \\
		
		\hline
		
		\caption{Symbols and values (if applicable) used in the slurry model. Group 1: fundamental thermodynamic variables. Group 2: slurry notation. Group 3: Dimensionless numbers. Group 4: Physical constants. Group 5: Seismic properties. Group 6: Properties determined from \textit{ab initio} calculations and high-pressure experiments. Group 7: Properties that assume ideal solution theory. Group 8: Heat flow estimates.}
		\label{tbl:val}
	\end{longtable}
	
	\section{Variations in the solid fraction, $\mathrm{d} \phi$} \label{sec:dphi}
\begin{figure}
		\centering
		\includegraphics[width=0.8\textwidth]{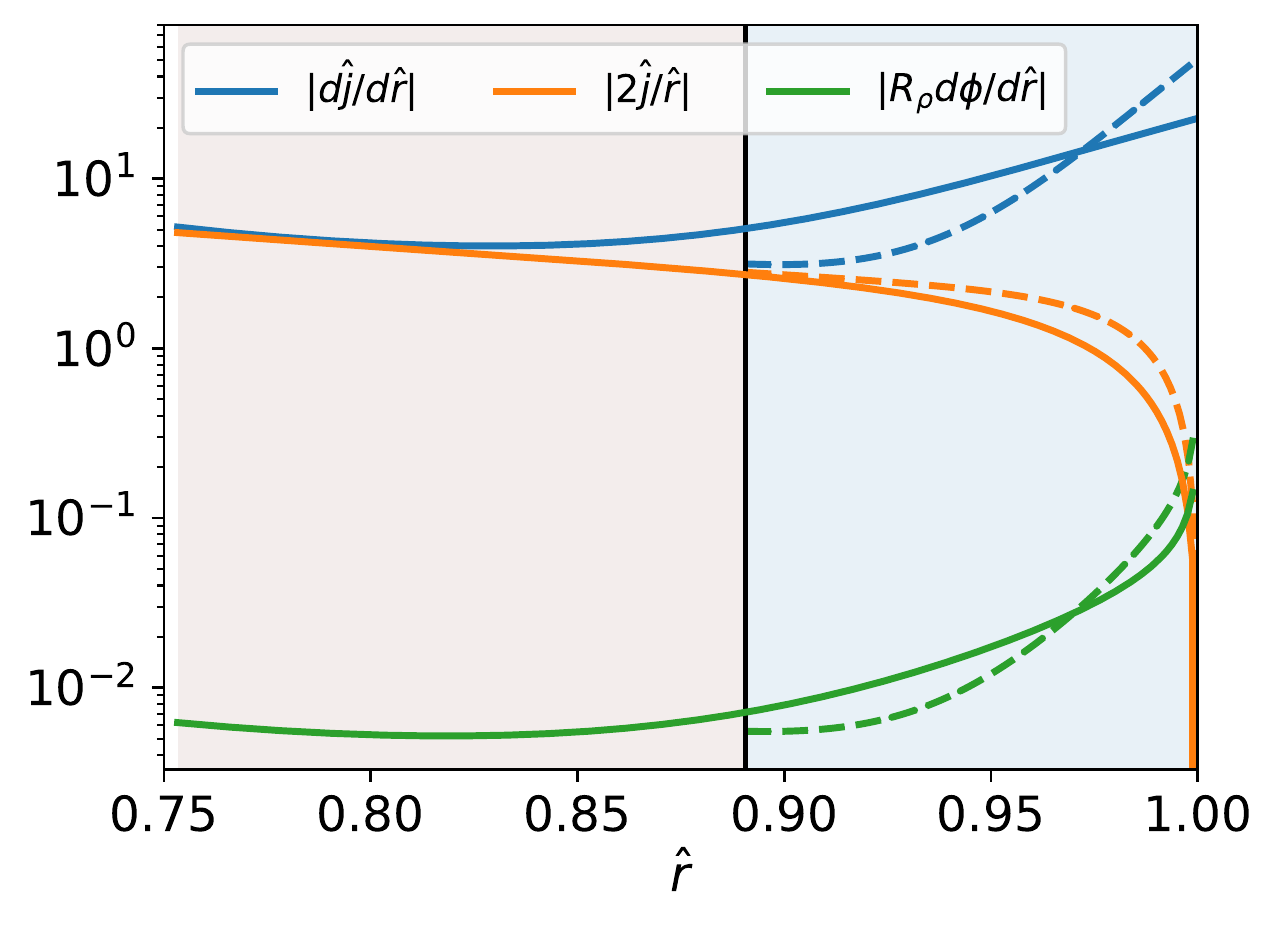}
		\caption{Comparing magnitudes $|\od{\hat{\jmath}}{\hat{r}} |$, $|\frac{2}{\hat{r}}\hat{\jmath}|$ and $|R_\rho \od{\phi}{\hat{r}}|$, across the slurry for layer thicknesses of $150$ (dashed, blue fill) and $400 \ \mathrm{km}$ (solid). Control parameters are $Q^s = 2.5 \ \mathrm{TW}$, $Q^{sl}= 5.0 \ \mathrm{TW}$ and $k = 100 \ \mathrm{W} \mathrm{m}^{-1} \mathrm{K}^{-1}$ (colour online).}
		\label{fig:dphi_ene}
	\end{figure}

	We suppose that the variations in the solid fraction, $\mathrm{d} \phi$, are negligible. Here we assess the validity of this assumption \textit{a posteriori}. The variation in solid fraction enters the energy equation (\ref{eqn:ene}) and the equation of state (\ref{eqn:den_flu}). In the first instance, including these variations adds an extra term to the latent heat release due to phase change. The dimensionless energy equation (\ref{eqn:nTemp}) becomes
		\begin{align*}
			-\hat{v} \od{\hat{T}}{\hat{r}} &= \frac{Le}{Pe} \left(\odtwo{\hat{T}}{\hat{r}} + \frac{2}{\hat{r}}\od{\hat{T}}{\hat{r}} \right) + \frac{1}{St} \left(\od{\hat{\jmath}}{\hat{r}} 	+\frac{2}{\hat{r}}\hat{\jmath} - R_\rho \od{\phi}{\hat{r}}\right).
		\end{align*}		
	Using the control parameters from Figure \ref{fig:eos}, we compare the order of magnitude of each part of the last term in the expression above where the $\mathrm{d} \phi / \mathrm{d} \hat{r}$ term appears. In Figure \ref{fig:dphi_ene} we can see that with $r^i/r^{sl} = 0.89$ and $0.75$ (corresponding to $d = 150$ and $400 \ \mathrm{km}$, respectively), the extra term due to the variations in $\phi$ is almost three orders of magnitude smaller compared with the other two terms in the majority of the layer. We observe that $|\frac{2}{\hat{r}}\hat{\jmath}| \sim  |R_\rho \od{\phi}{\hat{r}}|$ approaching the top of the layer, which is a consequence of $\hat{\jmath}$ tending to $0$ as dictated by boundary condition $\hat{\jmath}(1) =0$ in equation (\ref{eqn:nondim_jCSB}). This occurs within a very thin region at the top of the layer and so $|R_\rho \od{\phi}{\hat{r}}|$ makes a negligible difference to the overall energy balance in the slurry.
	
	Turning to the equation of state we observe that the contributions from the variations in solid also arise by imposing boundary condition (\ref{eqn:nondim_jCSB}), which can be evaluated analytically by assuming Stokes' flow and writing
		\begin{align*}
			\lim_{\hat{r} \to 1} \left( \od{\phi}{\hat{r}} \right) = \lim_{\hat{r} \to 1} \left( - \frac{3}{5} \left(\frac{\rho_-^s v_f}{K_\phi} \right)^\frac{3}{5} (- \hat{\jmath})^{-\frac{2}{5}} \od{\hat{\jmath}}{\hat{r}} \right) = - \infty
		\end{align*}
	On a physical basis Stokes' flow is unlikely to hold in the turbulent mixing region close to the CSB, and the contributions from the variations in $\phi$ in this thin region are considered to be a numerical artefact of maintaining $\hat{\jmath}(1) =0$. This thin region comprises less than 5\% of the layer and has a limited impact on assessing the overall stability of the slurry.
	
	\section{CMB heat flow, $Q^c$} \label{sec:cmb}
	In general contributions to the CMB heat flux can be separated into the secular cooling, $Q_s$, gravitational power, $Q_g$, and the latent heat $Q_l$, while pressure freezing is neglected and radiogenic heating is ignored for the sake of simplicity \cite{Gubbins04b}. This gives a core heat balance of
	\begin{align}
		Q^c = Q_s^l + Q_g^l + Q^{sl},
	\end{align}
	where
	\begin{align}
		Q_s^l = - \frac{c_p}{T_c} \od{T_c}{t} \int_{V^l} \rho T_a \ \mathrm{d}V^l \label{eqn:sec}
	\end{align}
	is the secular cooling in the liquid volume, $V^l$, and
	\begin{align}
		Q^l_g = \int_{V^l} \rho \psi \alpha_\xi \od{\xi}{t} \ \mathrm{d} V^l - 4 \pi (r^{sl})^2 \rho^{sl} \psi^{sl} \alpha_\xi \xi^{sl} v \label{eqn:grav}
	\end{align}
	is the gravitational power in the liquid volume, and $Q^{sl}$ is the heat flux imposed at the CSB that also contains $Q_l$. In (\ref{eqn:sec}), the adiabatic temperature is given by \cite{Gubbins03}
	\begin{align}
		T_a(r) = T^{sl} \exp \left(- \int_{r^{sl}}^{r} \frac{g \gamma}{\phi} \ \mathrm{d}r \right ).
	\end{align}
	We calculate the core cooling rate, $\mathrm{d}T_c / \mathrm{d}t$, by constructing an adiabat anchored at the present day CSB temperature, followed by constructing a new adiabat anchored at a new CSB temperature after time $\Delta t$ due to the advancing layer, and then finding the resulting decrease in the CMB temperature, $T_c = T_a(r_c)$. This is given by
		\begin{align*}
			\od{T_c}{t} \approx \frac{\Delta T_c}{\Delta t} = \frac{T^{sl}(r^{sl}+v \Delta t) \exp\left(- \int_{r^{sl}+v \Delta t}^{r_c} \frac{g \gamma}{\phi} \ \mathrm{d}r \right)  - T^{sl}(r^{sl}) \exp \left(- \int_{r^{sl}}^{r_c} \frac{g \gamma}{\phi} \ \mathrm{d}r \right)}{\Delta t}.
		\end{align*}
	The gravitational power (\ref{eqn:grav}), is composed of two parts: the first part is proportional to the change in oxygen concentration in the bulk of the volume, $V^l$, and the second part is from the motion of the CSB, where there is no additional contribution from the motion of the ICB since the CSB and ICB move at the same rate. By mass conservation, the change in oxygen concentration in the bulk is given by
		\begin{align*}
			\int_{V^l} \pd{\xi}{t} \ \mathrm{d} V^l = \frac{- 4 \pi (r^{sl})^2 \rho^{sl} \xi^{sl} v}{M_l},
		\end{align*}
	where $M_l = \int_{V^l} \rho \ \mathrm{d}V^l$ is the mass of the liquid outer core.
	
	\section{Definition of the solid expansion coefficient, $\alpha_\phi$} \label{sec:def}
	The coefficient, $\alpha_\phi$, is a dimensionless expansion coefficient that influences the contribution of the solid fraction to the slurry density. From equation (A3) of W18, we have that
	\begin{align}
		\alpha_\phi \equiv - \rho ^{sl} \left( \pd{V}{\phi} \right)_{p,T,\xi}. \label{eqn:app_phi}
	\end{align}
	From (A1) of W18, the expression for the Gibbs free energy, $d\Phi$, defines the specific volume as
	\begin{align}
		V \equiv \left(\pd{\Phi}{p} \right)_{T,\xi,\phi}. \label{eqn:app_vol}
	\end{align}
	The lever rule in equation (A9) of W18 gives
	\begin{align}
		\left(\pd{\Phi}{\phi} \right)_{p,T,\xi} = \Phi^s - \Phi^l \label{eqn:app_lev}
	\end{align}
	and
	\begin{align}
		\left(\pd{\Phi^s}{p} \right)_{T,\xi,\phi} = V\sFe, \hspace{1cm} \left(\pd{\Phi^l}{p} \right)_{T,\xi,\phi} = V\lFeO, \label{eqn:app_vol2}
	\end{align}
	where $\Phi^s$ is the solid part of the Gibbs free energy and $\Phi^l$ is the liquid part. Substituting (\ref{eqn:app_vol}), (\ref{eqn:app_lev}) and (\ref{eqn:app_vol2}) into (\ref{eqn:app_phi}) gives the final result
	\begin{align}
		\alpha_\phi
		% &= - \rho ^{sl}  \pd{}{\phi} \left( \pd{\Phi}{p} \right)_{p,T,\xi} \\
		% &= - \rho ^{sl} \pd{}{p} \left( \pd{\Phi}{\phi} \right)_{p,T,\xi} \\
		% &= - \rho ^{sl} \pd{}{p} \left(\Phi^s-\Phi^l \right) \\
		&= - \rho ^{sl} \left(V\sFe - V\lFeO \right) = \rho ^{sl} \Delta V \slFeO.
	\end{align}

\end{document}